\documentclass[aps,pre,twocolumn,showpacs,superscriptaddress]{revtex4-2}
\usepackage{graphicx}  % needed for figures
\usepackage{dcolumn}   % needed for some tables
\usepackage{bm}        % for math
\usepackage{dsfont}    % for math (double-stroked characters via \mathds{text})
\usepackage{amssymb}   % for math
\usepackage{amsmath}
\usepackage{journals}
\usepackage{color}
\usepackage{siunitx}
\usepackage{physics}
\usepackage{hyperref}

\usepackage{booktabs}

\ProvideTextCommand{\DJ}{OT1}{\leavevmode\raisebox{-.5ex}{\makebox[0pt][l]{\hskip-.07em\accent"16\hss}}D}

\usepackage[utf8]{inputenc}

\begin{document}
\title{Learning Dynamic Graphs, Too Slow}
\author{Andrei A.~Klishin}
\affiliation{Department of Bioengineering, University of Pennsylvania, 
	Philadelphia, PA 19104 USA}
\author{Nicolas H.~Christianson}
\affiliation{Department of Computing and Mathematical Sciences, California Institute of Technology, Pasadena, CA 91125 USA}
\author{Cynthia S.~Q.~Siew}
\affiliation{Department of Psychology, National University of Singapore, Singapore 117570 Singapore}
\author{Dani S.~Bassett}
\email{dsb@seas.upenn.edu}
\affiliation{Department of Bioengineering, University of Pennsylvania, 
	Philadelphia, PA 19104 USA}
\affiliation{Department of Physics \& Astronomy, University of Pennsylvania, 
	Philadelphia, PA 19104 USA}
\affiliation{Department of Electrical \& Systems Engineering, University of 
	Pennsylvania, Philadelphia, PA 
	19104 USA}
\affiliation{Department of Neurology, University 
	of Pennsylvania, Philadelphia, PA 19104 USA}
\affiliation{Department of Psychiatry, University 
	of Pennsylvania, Philadelphia, PA 19104 USA}
\affiliation{Santa Fe Institute, Santa Fe, NM 87501 USA}

\date{\today}

\begin{abstract}
	The structure of knowledge is commonly described as a network of key 
	concepts and semantic relations between them. A learner of a particular domain can discover this network by navigating the nodes and edges presented by instructional material, such as a textbook, workbook, or other text. While over a long temporal period such exploration processes are certain to discover the whole connected network, little is known about how the learning is affected by the dual pressures of finite study time and human mental errors. Here we model the learning of linear algebra textbooks with finite length random walks over the corresponding semantic networks. We show that if a learner does not keep up with the pace of material presentation, the learning can be an order of magnitude worse than it is in the asymptotic limit. Further, we find that this loss is compounded by three types of mental errors: forgetting, shuffling, and reinforcement. Broadly, our study informs the design of teaching materials from both structural and temporal perspectives.
\end{abstract}

\maketitle

\section{Introduction}

Knowledge structures built by humans are best described as networks \cite{schapiro2013neural, karuza2016local, engelthaler2017feature, sizemore2018knowledge, solomon2019implementing, peer2021structuring}. Such networks can be built when acquiring one's first language \cite{Saffran_Aslin_Newport_1996, romberg_saffran_2010, stella2017multiplex}, engaging in curiosity-driven free exploration \cite{lydon2021hunters}, or accumulating scientific knowledge throughout humanity's history \cite{ju2020network}. Network representation emphasizes not just a listing of known facts (nodes), but also the intricate pattern of interconnections between these facts (edges) \cite{zurn2021edgework}. These connections also strongly affect how the network evolves, from discovering brand new nodes and edges \cite{iacopini2018network} to learning them from an external source such as a formal class \cite{siew2019using}.

The structure of networks is easily revealed by dynamical processes upon them, such as random walks \cite{costa2006learning,costa2007exploring} or more complex exploration algorithms \cite{asztalos2010network}. Unconstrained memory processes such as random walks mimic several of the features of spontaneous thought \cite{mildner2019spontaneous}. Random walk measures also capture the indirect relationships among words in semantic networks better than distributional measures of language corpora \cite{de_deyne_predicting_2016}. Controlled psychological experiments show that humans intuitively and efficiently learn the structure of networks from the statistics of a sequence of stimuli \cite{lynn2020review, karuza_value_2022}. This learning is imperfect: allowing for mental errors reduces the cognitive load on the brain but also highlights the higher-order structure of networks at the expense of fine detail \cite{michaelian2011epistemology, kahn2018network, lynn2020errors, zeng2021tracking}. This optimization of cognitive effort in the brain is hypothesized to shape the communication network architectures themselves \cite{ramonicancho2003least, zurn2020network, lynn2020human}, and has an established neurophysiological basis \cite{momennejad2017successor, stachenfeld2017hippocampus, Stiso2021, ryan2022forgetting}.

Learning can be studied at two scales: that of large networks which encompass the full knowledge of an individual or a society, and that of small synthetic networks which are used in memory experiments. Yet typical educational scenarios lie in between these two extremes: a learner is expected to acquire a finite number of connected concepts from a finite lecture course or a single textbook \cite{cramer2018network, siew2020applications}, each of which has its own particular network structure \cite{yun2018extraction, christianson2020architecture, vukic2020structural}. The success of learning is measured not by memorizing the text of the educational medium, but by understanding key concepts and their interconnections \cite{freeman2014active, denervaud2021education} which support problem-solving \cite{corbett2010cognitive, palazzo2010patterns}. At the same time, exposure to the same educational materials can result in building very different networks, owing both to varying learner effort and memory imperfections \cite{koponen2018concept, siew2019using, nilsson2021structural}. Learners also vary in how they identify the key concepts \cite{lommi2019landmarks} and the genealogical relations between them \cite{koponen2018genealogy}.

The learning sciences have had an enduring focus on learning mechanisms that lead to better memory of studied materials (see \cite{weinstein_teaching_2018} for an overview). For example, the effects of retrieval practice, where learners who practice retrieving information from memory do better at a test than learners who simply re-studied the material \cite{roediger_test-enhanced_2006}, are well-established, as are the effects of interleaved practice, where better learning occurs when different to-be-learned skills or topics are continuously alternated \cite{kang_learning_2012}. In contrast, relatively less attention has been directed to the nature of the \emph{inputs} for learning. This lack is striking given that (i) a core principle of learning predicts that the frequency of input shapes what is being learned \cite{estes_statistical_1955}, and (ii) human learners are highly sensitive to statistical associations encountered in the environment, even through passive or incidental exposure \cite{Saffran_Aslin_Newport_1996, karuza_value_2022}. Ref.~\cite{braithwaite_children_2018} demonstrated that students pick up spurious correlations between features of problem sets and solution strategies in mathematics textbooks. This behavior could have implications for mathematics learning because students use these features to decide on a solution strategy even when the feature is irrelevant. Given that there are observable effects of input set characteristics on learning \cite{braithwaite_computational_2017}, techniques are needed to represent the structure of the learning environment \cite{christianson2020architecture} and also to model the mapping of this structure to the mental models that learners acquire through their interactions with the learning environment.

In this paper we trace how learner efforts and memory effects drive the formation of learned mental models from taught semantic networks. For the example teaching material, we use networks of key concepts extracted from ten popular linear algebra textbooks \cite{christianson2020architecture}. As the learner progresses through the text, the network structure afforded to them by the book evolves as well. We model learning as a random walk on this temporal network and study the statistics of learning through a combination of numerical stochastic simulations and analytical exposure theory as introduced in Ref.~\cite{klishin2022exposure} and expanded upon here. We find that the network learned with finite effort is significantly incomplete, containing fewer network nodes and edges, and that this effect is further compounded by memory imperfections. One particular form of mental error, random shuffling of the order of stimuli, mostly leads to worse learning of already presented network edges but can afford a limited ability to anticipate future edges. This work offers a principled way to predict the learned network structures from taught ones, for a variety of learners. In educational settings, this prediction can serve as an important forewarning and thus be used to tweak the instructional materials to emphasize some concepts and connections over others to ensure that they are learned.

\section{Learning model}
\begin{figure*}
	\includegraphics[width=0.8\textwidth]{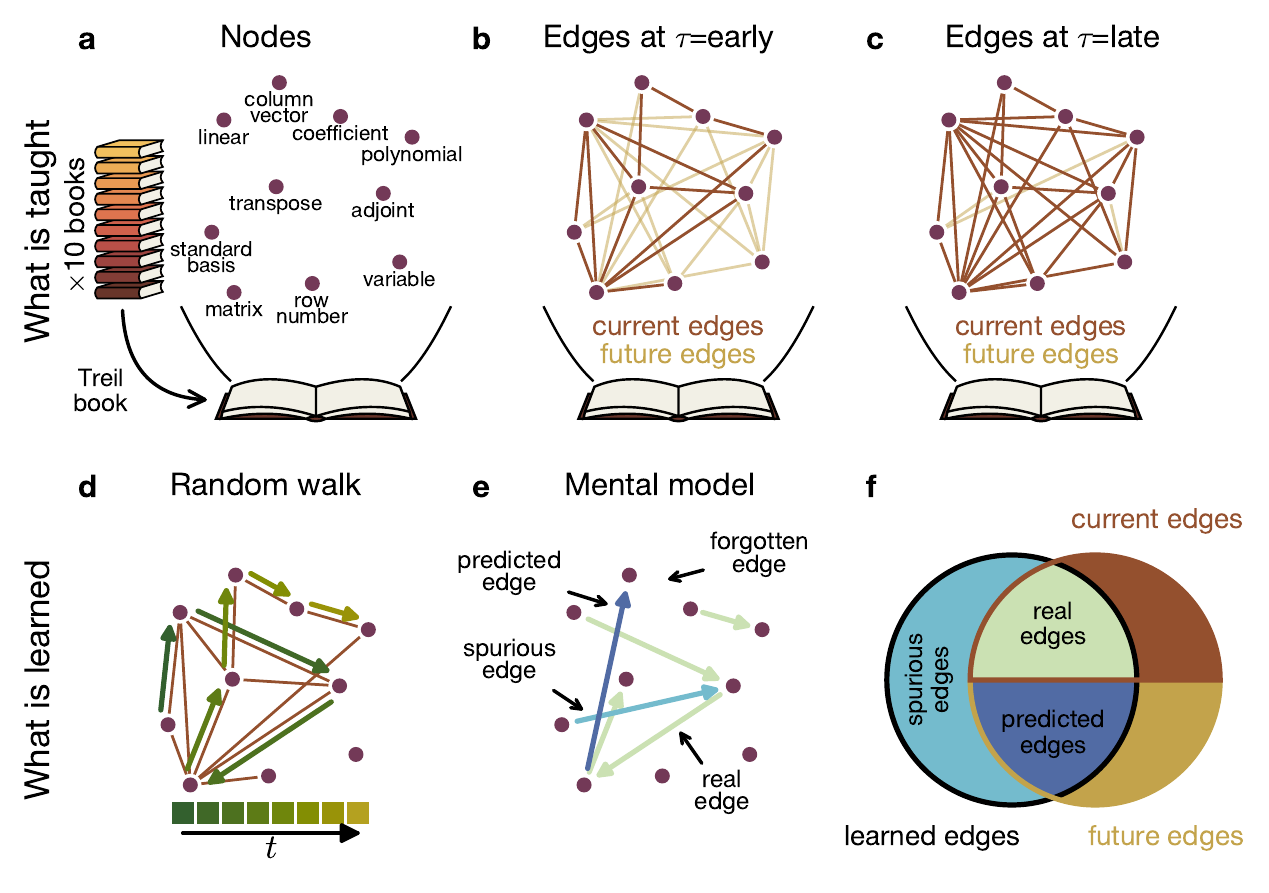}
	\caption{\textbf{Linear algebra textbooks provide a substrate for learning 
	a semantic network of concepts.} (a) Within each textbook, important linear algebra concepts serve as the nodes of the semantic network. (b-c) Co-occurrences of concepts within the same sentence form the edges of the semantic network. By a given sentence $\tau$ of the book, some edges have already been established (brown), while others appear later (yellow). (d) The learning process consists of random walk steps along the edges that already exist. (e) The mental model of the network is generated by an imperfect human memory process: some edges are remembered correctly, while others are forgotten or misplaced. (f) The set of learned edges overlaps imperfectly with the taught current and future edges, as shown in this Venn diagram.}
	\label{fig:setup}
\end{figure*}

\subsection{What is taught}
In this paper we are interested in a human subject learning a structured 
semantic network of concepts. This is a very common scenario of learning from instructional material, such as a textbook, a lecture, or a webpage. 
Learning from a single fixed source is a building block of a more 
open-ended, self-directed learning process that draws on multiple sources, 
synthesizes the information, and relays it to others.

In order to obtain quantitative insights into the learning process, we must 
establish models of two complementary processes: how information is 
\emph{taught} by the source material and how it is \emph{learned} by the human 
(Fig.~\ref{fig:setup}). Notably, the goal of such learning is not to reproduce 
the material verbatim (as would be necessary for reciting a poem or a musical 
piece), but to reconstruct the pattern of connections between the concepts that can be dynamically traversed. On this view, the source material exposes a 
range of concepts (Fig.~\ref{fig:setup}a) and their connections of varying strength, which might also 
change in time (Fig.~\ref{fig:setup}b-c). The learner takes a random walk along the conceptual connections and records a memory of the sequence of steps (Fig.~\ref{fig:setup}d). 
The resulting memory is a learned, or reconstructed, semantic network that 
sometimes closely follows the source material, and other times differs from it in important ways (Fig.~\ref{fig:setup}e). Qualitatively, the relationship between the taught and the learned networks can be drawn as a Venn diagram (Fig.~\ref{fig:setup}f). In order to make this relationship quantitative, below we outline the mathematical model of learning.

We demonstrate our model of teaching and learning on a concrete set of ten 
networks extracted from popular linear algebra textbooks in 
Ref.~\cite{christianson2020architecture}. For each textbook, the authors 
identified the set of important mathematical concepts such as ``matrix'' or 
``polynomial''. Two concepts $i,j$ are deemed connected if they are mentioned 
together in a sentence. The number of such co-occurrences is given by the elements $A_{ij}$ of the weighted symmetric adjacency matrix, whereas the sentence of the first co-occurrence is given by the elements $F_{ij}$ of the filtration matrix  (see Appendix~\ref{app:textbooks} for a comparison of basic textbook statistics). Notably, the textbook network provides a dynamic substrate for random walks. The probability of going from node $i$ to 
node $j$ is given by:
\begin{align}
	P(j|i)=T_{ij}(\tau)=\frac{A_{ij}(\tau)}{\sum_j A_{ij}(\tau)};\; 
	A_{ij}(\tau)=A_{ij}\cdot [F_{ij}<\tau],
	\label{eqn:Pji}
\end{align}
where $T_{ij}(\tau)$ are the elements of the row-normalized transition matrix, $\tau$ indexes progress through the book as measured by sentences, and $[\cdot]$ is an indicator function equal to 1 if the expression inside is true, and 0 otherwise. Throughout the paper, we use $A_{ij}$ without an argument to refer to the number of co-occurrences and $A_{ij}(\tau)$ with an argument as an explicitly temporal element of the adjacency matrix. In other words, each edge $(i,j)$ of the network appears at full strength at the time point given by its filtration order $F_{ij}$. The random walk is thus restricted to walking along the connections that have already been introduced. For the textbooks we consider, all newly introduced nodes attach to the main connected component of the network within just a few sentences, so the network stays connected.

Every learner is exposed to the same temporal order of network evolution, yet 
not all learners put equal effort towards learning. One can skim the textbook 
quickly, or comb through every line and re-derive every proof and exercise. The
amount of effort students put into learning might not be susceptible to nudging
\cite{oreopoulos2019remarkable}, but can vary depending on interest and 
curiosity \cite{peterson2019case}. Mathematically, we model this varying effort with the variable of \emph{dilation} $D\equiv t/\tau$, where $t$ is the learning time measured in random walk steps and $\tau$ is the evolution time measured in sentences. Dilation is thus the average number of random walk steps per sentence, which might be much smaller than 1 (skimming the text) or much larger than 1 (studying thoroughly). As the dilation gets arbitrarily large $D\to \infty$, the learner samples every possible transition in the network. In the investigations below we treat dilation $D$ as a free parameter.

\subsection{What is learned}
In order to keep track of the built mental model of the network, we introduce 
the integer-valued \emph{memory matrix} $\mathbf{M}$ where the indices run over the network nodes. At the beginning of the random walk, all entries $M_{ij}$ of the matrix are set to zero, and then are incremented as the random walk advances. Typically, when the learner observes the transition $i\to j$, one count is added to the entry $M_{ij}$. However, because of certain human memory effects that we discuss below, the count can be misplaced or removed altogether from the memory matrix.

Once the memory matrix is obtained, we can construct the empirical transition matrix by performing a row normalization element-wise:
\begin{equation}
	\hat{T}_{ij}=
\begin{cases}
		\frac{M_{ij}}{\sum_j M_{ij}},& \sum_j M_{ij}\neq 0\\
		0,& \text{otherwise}
\end{cases},
\label{eqn:mentalmodel}
\end{equation}
where the second case reflects the fact that if the learner remembers no transitions out of node $i$, they are unable to estimate the transition probabilities. Since the matrices of the textbook network are sparse, the memory matrix is also typically sparse. Note that the absolute number of memory
counts cancels out from the expression; only the relative populations of the entries matter.

We assess network learning by comparing the empirical transition matrix 
$\hat{\mathbf{T}}$, the instantaneous transition matrix of the textbook 
$\mathbf{T}(t)$, and the final transition matrix of the full textbook 
$\mathbf{T}$, which represents the complete network that the student could hypothetically learn. The three matrices have partial overlaps (Fig.~\ref{fig:setup}f): not all taught edges might be learned and not all learned edges might have been taught. In order to compare the matrices, we define a set of metrics such as precision and recall, which treat one of the matrices as a set of real numbers and another as a binary filter. In addition to the memories of edges, we can compute the memories of visitation of nodes by performing column-summation $\sum_j M_{ij}$, for which metrics can be defined similarly. The goal of our investigation is to understand how the learning metrics depend on the progress through the book $\tau$, the dilation $D$, and the memory effects described below. The mathematical details of these 
whole-network metrics are given in Appendix~\ref{app:precrec}.

\subsection{Memory effects}
\begin{figure}
	\includegraphics[width=\columnwidth]{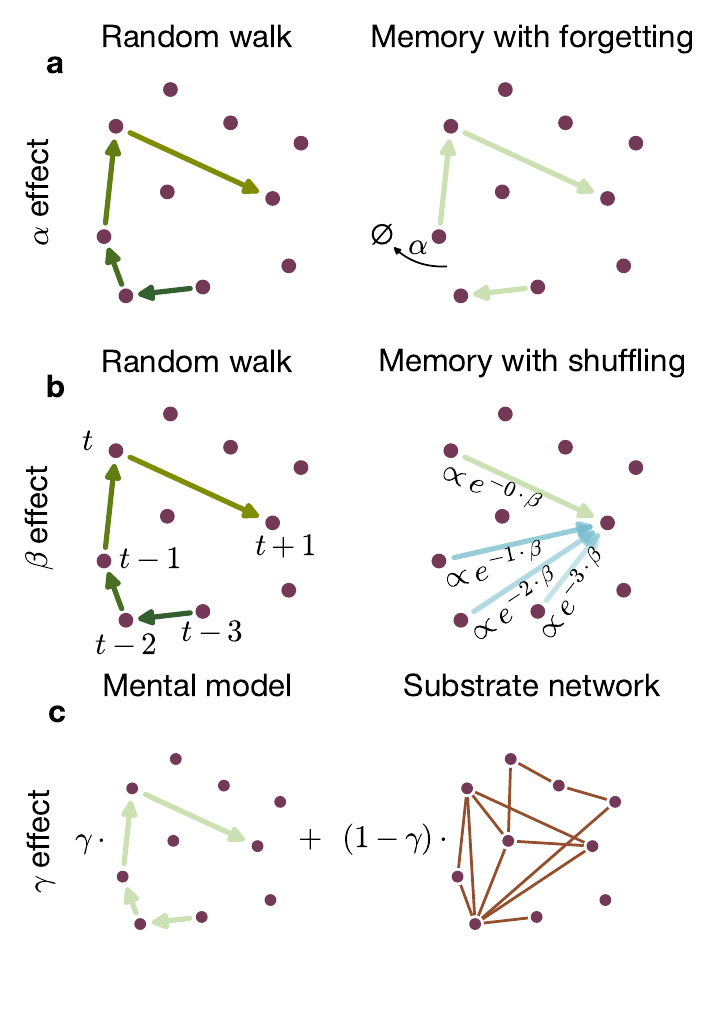}
	\caption{\textbf{Memory effects shape the learning of networks by humans.} (a) The $\alpha$ effect corresponds to random forgetting of learned network edges with probability $\alpha$ per random walk step. (b) The $\beta$ effect corresponds to random shuffling of memories: the destination of a random walk step is always the correct node $t+1$ but the origin is confused between $t$, $t-1$, $t-2$ etc., with decaying probability. (c) The $\gamma$ effect corresponds to the reinforcement of memories: the random walker mixes the existing memories of weight $\gamma$ with the substrate network of weight $(1-\gamma)$.}
	\label{fig:effects}
\end{figure}

Human memory is an associative, distributed psychological function, the purpose of which is not to provide a perfect replica of past events, but to extract the gist, produce abstractions and inferences form the observed data, and create an overall generalizable model of the world \cite{michaelian2011epistemology, lynn2020human, zeng2021tracking}. In order to mimic the features of this model-making, we consider memory acquisition to be a digital process of adding integer memory counts, modified by three different effects which we parameterize with $\alpha$, $\beta$, and $\gamma$ (precise definitions given later). The presence of each of these effects has separately been confirmed in human behavioral experiments, but generically we expect them to be present all at the same time. Each of the effects is known to vary in magnitude, which justifies studying a range of $\alpha, \beta, \gamma$ parameter values.

While the three effects can have similar bearing on the learning metrics, they act through orthogonal mechanisms (Fig.~\ref{fig:effects}). The $\alpha$ effect corresponds to forgetting of encountered network edges or conceptual transitions (Fig.~\ref{fig:effects}a) \cite{wozniak1995two, rubin1996one, rubin1999precise}. The $\beta$ effect corresponds to a temporal shuffling or random reordering of encountered transitions in memory (Fig.~\ref{fig:effects}b) \cite{lynn2020errors}. The $\gamma$ effect corresponds to reinforcing the random walk by the memories of prior transitions (Fig.~\ref{fig:effects}c) \cite{iacopini2018network, lydon2021hunters}. All three effects can be present at the same time and interact with each other. While other memory effects can exist as well, focusing on these three both establishes an important intuition for mental model distortion and develops transferable mathematical techniques. Importantly, the effects are heavily compounded by the finite length of any learner's random walk (finite dilation $D$), which leads to a marked under-sampling of the nodes and edges.

\subsection{Mathematical formalisms}
In order to disentangle these three effects and the effect of under-sampling, we utilize three mathematical lenses of analyses. First, we use a mean field theory that assumes a fully equilibrated random walk on the network and that captures long-time network statistics and distortions, akin to Refs.~\cite{lynn2020errors,lynn2020human}, but cannot account for under-sampling effects. Second, we perform direct stochastic simulations of random walks on dynamic networks, which sample both the random walk steps and the memory effect realizations (details of the simulations are given in Appendix~\ref{app:simulation}). Third, we perform computations using exposure theory, as derived and validated in Ref.~\cite{klishin2022exposure}, which gives closed-form approximations for under-sampling effects (a primer on exposure theory is given in Appendix~\ref{app:exposure}).

The goal of exposure theory is to provide analytic expressions for the probability distributions of memory counts $M_{ij}$. Under certain approximations, these distributions take Poisson shape parameterized by a single exposure value $E_{ij}(t)$ for each edge. The Poisson distributions of memory counts and the rules for event aggregation define an effective statistical mechanics framework of random processes on complex networks. Tracking the deterministic evolution of the exposure value offers a speed-up by many orders of magnitude of computational time. While the original formulation of exposure theory in Ref.~\cite{klishin2022exposure} assumed that each random walk step follows the substrate network and is remembered correctly and in perpetuity, here we derive extensions that account for the effects of forgetting, shuffling, and reinforcement (see Appendix~\ref{app:exposure}).

\section{Dilation and under-sampling}
\subsection{Global exploration}
\begin{figure}[t]
	\includegraphics[width=\columnwidth]{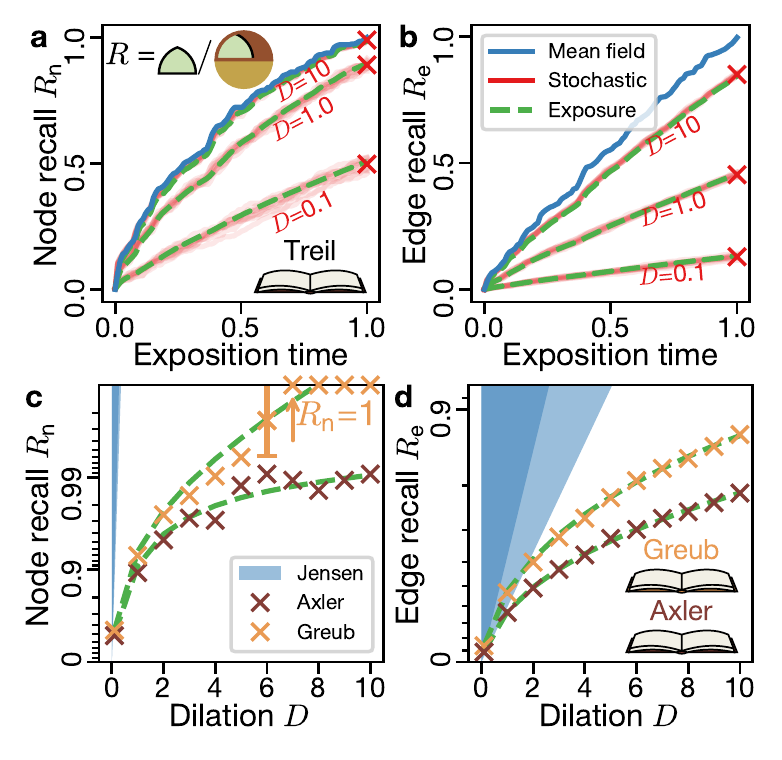}
	\caption{\textbf{Finite-time learning leads to incomplete node and 
	edge exploration in textbook networks.} (a-b) Node and edge recall $R_\text{n},R_\text{e}$ of the random walker on the Treil textbook network. Red curves show 10 replicas of stochastic simulations at each dilation; green dashed curves show the exposure prediction. At higher dilation $D$ both recall curves follow the mean field curve (blue) more closely than at lower dilation. (c-d) Node and edge recall by the end of the two textbooks, asymptotically approaching 1. For the Greub textbook, node recall discontinuously jumps to $R_\text{n}=1$ at finite $D$. The blue shading shows the Jensen bound on learning speed, its straight boundary on a semi-log plot corresponds to an exponential approach to $R=1$.}
	\label{fig:alpha1}
\end{figure}

First we consider the under-sampling of the network both along the exposition time of the book and by the end of it. The random walk samples from the nodes and edges present in the network at a given time. The fraction of nodes and edges learned can be measured by the recall metrics $R_\textsf{n}$ and $R_\textsf{e}$, or the fraction of learned nodes and edges with respect to all current and future ones (Fig.~\ref{fig:alpha1}a inset). As established previously \cite{christianson2020architecture}, the nodes and edges are introduced sublinearly throughout the text: faster initially, and slower by the end. This introduction rate then sets the upper limit on exploration by a mean-field random walk (Fig.~\ref{fig:alpha1}a,b).

The degree to which recall follows the mean-field limit depends on the
dilation $D$, or the length of the random walk that the learner takes upon the network. We contrast the number of nodes and edges recalled by high versus low $D$ learners to model extensive versus light study habits. For high values of dilation $D=10$ the stochastic trajectories of $R_\textsf{n}$ lie just below the mean field one, such that any newly introduced node is quickly discovered (Fig.~\ref{fig:alpha1}a). In contrast, for low values of dilation $D=0.1$ the stochastic trajectories lag far behind the mean field: barely half of the nodes are discovered by the end of the book. Turning from recall of nodes to recall of edges, we find that stochastic trajectories fall even further behind the mean field predictions (Fig.~\ref{fig:alpha1}b): every random walk step can discover at most one node and one edge, but there are many more edges than nodes and thus edge learning is slower 
\cite{asztalos2010network}.

With growing dilation, mental models approach complete recall ($R_\textsf{n}=R_\textsf{e}=1$) for both nodes and edges, but the convergence is slow (Fig.~\ref{fig:alpha1}c-d). For the Greub textbook the node recall discontinuously jumps to $R_\textsf{n}=1$ near $D=6$, whereas for the Axler textbook we do not see this jump within the plotted range. These two distinct behaviors suggest that there exists some difference between the two textbook networks.
In Ref.~\citep{klishin2022exposure} we showed that average network learning speed is limited from above by the Jensen bound, which can only be saturated by unweighted networks. For all textbooks the stochastic and exposure recall grows much slower than allowed by the bound (Fig.~\ref{fig:alpha1}c-d and Appendix~\ref{app:supp}). Despite the efforts of textbook authors, the learning of nodes and edges is seemingly very inefficient when measured in the aggregate, and differs between textbooks. What causes the slowdown of learning and what drives the difference between the textbooks?

\subsection{Local learning of nodes and edges}
\begin{figure}
	\includegraphics[width=\columnwidth]{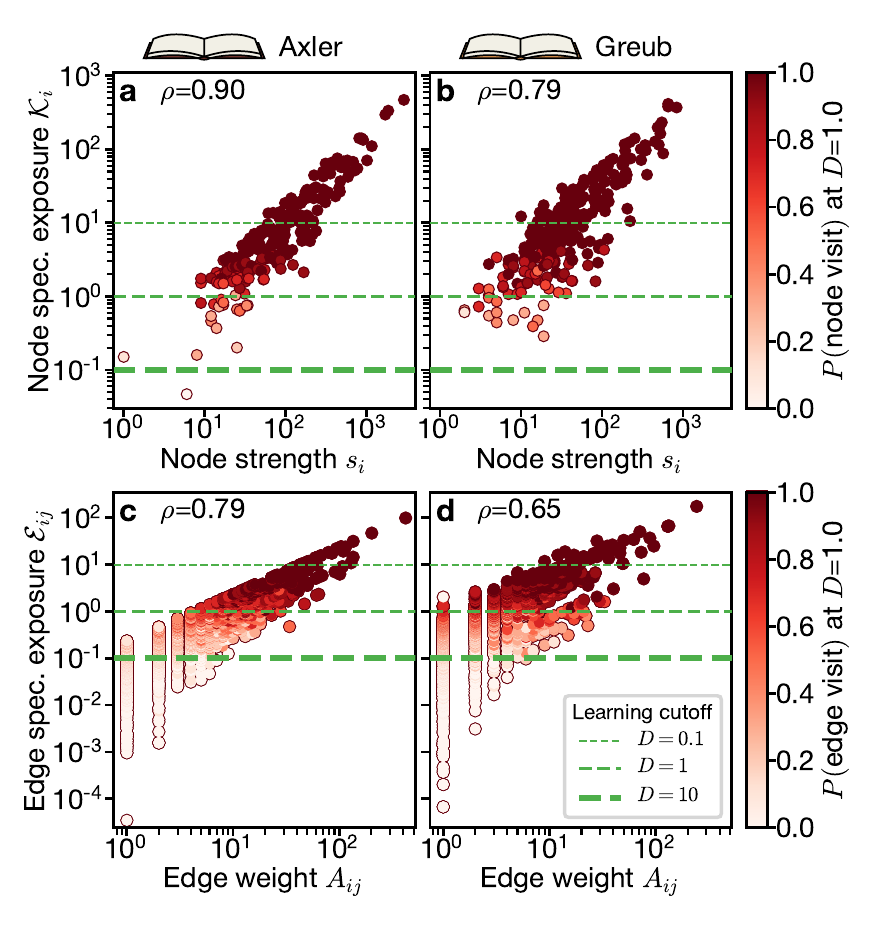}
	\caption{\textbf{Node and edge exposure predicts learning across orders of magnitude of dilation.} (a-b) Scatter plots of network nodes measured by specific exposure and strength. (c-d) Scatter plots of network edges measured by specific exposure and edge weight. Marker color corresponds to the probability of a node or edge learned across 10 stochastic replicas at $D=1.0$. Horizontal dashed lines indicate the boundary of node and edge learning at different values of the dilation $D$. Note that all of the Greub nodes lie above the $D=10$ line, but some of the Axler nodes lie below the $D=10$ line. In each panel $\rho$ is the Spearman correlation coefficient, $\log_{10}(p)<-12$, between the x- and y-axis variables.}
	\label{fig:alpha2}
\end{figure}

In order to explain the learning slowdown and its variance across textbooks, we consider the textbook network structure at a more granular level of individual nodes and edges. On one side, the final textbook network can be described by the static metrics of node strengths $s_i$ (weighted degrees) and edge weights $A_{ij}$. On the other side, the dynamics of both network growth and random walks on the network require dynamic metrics. In exposure theory, each node and edge \emph{deterministically} accumulates exposure, which is predictive of the \emph{stochastic} number of memories of that node or edge \cite{klishin2022exposure}. For nodes, at a given dilation $D$, the node \emph{integral} exposure by the end of the textbook is just $K_i=D\cdot \mathcal{K}_i$, where the node \emph{specific} exposure $\mathcal{K}_i$ depends on the network evolution trajectory but not the learner. Similarly for edges---the edge integral exposure is $E_{ij}=D\cdot \mathcal{E}_{ij}$, where $\mathcal{E}_{ij}$ is the edge specific exposure (see Appendix~\ref{app:exposure} for derivations). How do these static and dynamic metrics help us to understand granular network learning?

The network nodes span several orders of magnitude by exposure and some are reliably learned across many stochastic simulation runs while others are not (Fig.~\ref{fig:alpha2}a-b). Each node is learned with high probability for $K_i>1$, or, put differently, the threshold line $\mathcal{K}_i= 1/D$ serves as a \emph{linear classifier} separating the learned from the not learned nodes. As dilation gets higher, corresponding to more extensive study, the threshold moves lower, so that more and more nodes end up above the threshold and are learned. The distribution of nodes by specific exposure differs between the books: whereas all of Greub nodes (Fig.~\ref{fig:alpha2}b) are learned by $D=10$, not all of the Axler nodes are learned (Fig.~\ref{fig:alpha2}a). By comparison, the static metric of node strength $s_i$ is not as predictive of learning as the dynamic metric, even though it is strongly correlated with the node specific exposure. Exposure theory thus gives a very granular prediction of node learning.

We conduct a similar analysis for network edges, though there are many more edges than nodes, and edge specific exposure $\mathcal{E}_{ij}$ spans more orders of magnitude than node specific exposure $\mathcal{K}_i$. Just like nodes, edges are learned with high probability for $E_{ij}>1$ and thus the threshold line $\mathcal{E}_{ij}=1/D$ is a good linear classifier. Unlike for nodes, a dilation of $D=10$ is not sufficient to learn all of the edges. The difference is especially striking for the weakest connections $A_{ij}=\{1,2,3\}$, for which specific dilation ranges from $10^{-4}$ to $10^0$. If these very weak edges are introduced early in the textbook, then they can accumulate enough exposure to be learned. The later they are introduced, the less time they have to accumulate exposure, and the more other edges they compete with for exposure. While strong edges are likely to be discovered by all learners, most of the weak edges are unlikely to be discovered even by the most thorough learners.

The network heterogeneity by specific exposure of nodes and edges serves as a 
mechanism to \emph{prioritize} some concepts and connections over others to 
ensure that they are learned by all, even the most cursory of learners with low $D$. For static networks, the specific exposure of nodes is directly proportional to their strength $\mathcal{K}_i\propto s_i$ and the specific exposure of edges is strictly proportional to their weight $\mathcal{E}_{ij}\propto A_{ij}$ \cite{klishin2022exposure}; for dynamic networks like the ones presented here the specific exposure is a time integral of either node strength or edge weight and thus the strict proportionality reduces to a strong Spearman correlation ($\rho$ in Fig.~\ref{fig:alpha2}). In other words, the key mechanisms of prioritization are to mention a concept or a connection early on and repeat it frequently in the text.

We showed that globally network learning is much slower than allowed by the Jensen bound, but locally the network nodes and edges vary widely by priority of learning. Are the slowdown and the prioritization connected? From exposure theory we know that the average learning depends on the \emph{distribution} of nodes and edges by exposure rather than their absolute exposure values \cite{klishin2022exposure}. The Jensen bound for node learning is saturated when all nodes have the same specific exposure, while for edge learning it is saturated when all edges have the same specific exposure. For static networks, that scenario would correspond to regular and unweighted networks, respectively. Globally, such networks would have the fastest \emph{average} learning. However, locally such network learning would be very unpredictable: in regular networks all nodes are equally likely to be learned, and in unweighted networks all edges are equally likely. If independent learners sample from such networks, their mental models would not have the same priority and thus would be markedly different. In other words, network heterogeneity is simultaneously the cause of prioritization and slowdown, which can only appear together.

\section{The $\alpha$ effect}
\begin{figure}[t]
	\includegraphics[width=\columnwidth]{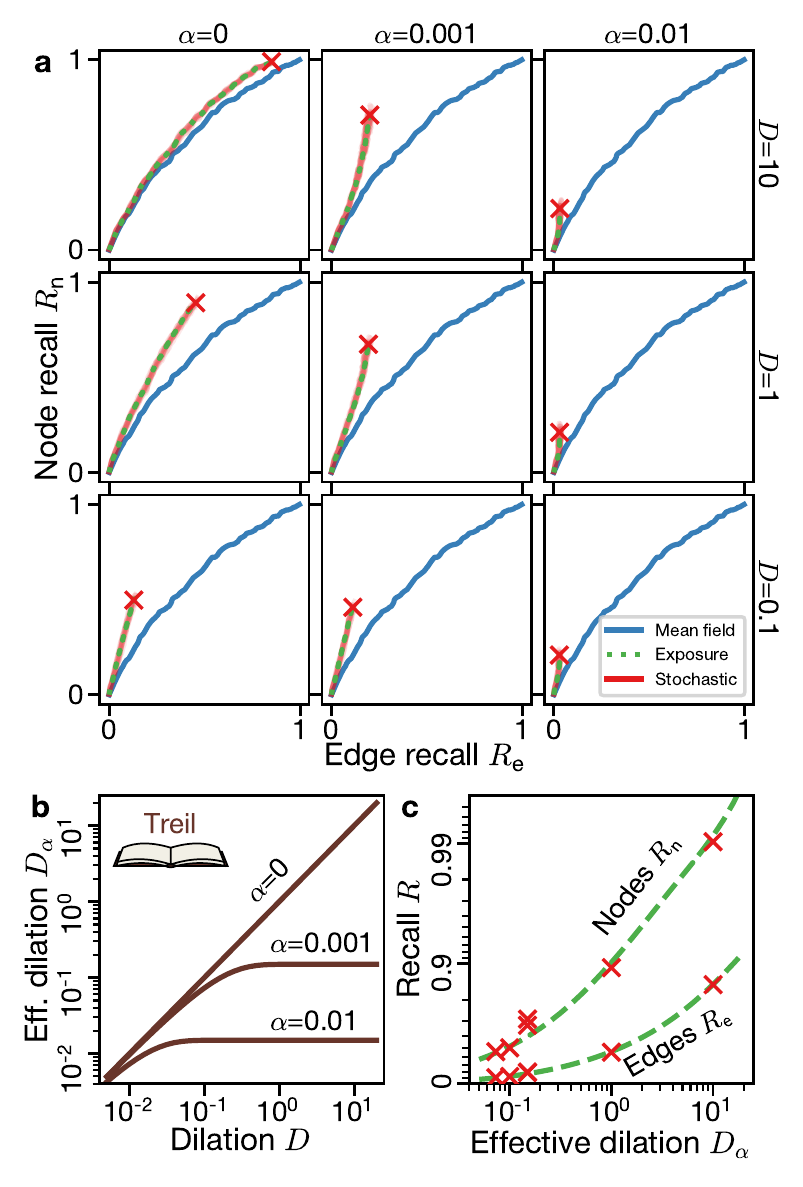}
	\caption{\textbf{Forgetting reduces the effective dilation and reduces learning.} (a) Learning trajectories 
	in $(R_\text{e},R_\text{n})$ space at varying dilation $D$ (rows) and 
	forgetting $\alpha$ (columns). The red cross marks the final position of 
	learning, averaged over 10 replicas. (b) Conversion of true dilation $D$ to effective dilation $D_\alpha$. At any finite forgetting $\alpha$ the effective dilation plateaus. (c) Effective dilation drives the data collapse of final recall of nodes and edges at varying parameter combinations $\alpha,D$. Each red cross corresponds to one of the parameter regimes from panel (a).}
	\label{fig:alpha3}
\end{figure}

Having established the basic intuition of under-sampling due to finite dilation, we now consider the $\alpha$ effect: stochastic forgetting. At each step of the random walk, precisely one new memory is added to the memory matrix $\mathbf{M}$. At the same time, every single memory has a small uniform chance $\alpha$ of being forgotten per step, which leads to an exponential distribution of memory lifetimes, consistent with empirical measurements of forgetting \cite{wozniak1995two, rubin1996one, rubin1999precise}. The forgetting caps the number of memories that the walker can hold at around $\sum_{ij}M_{ij}\simeq 1/\alpha$ with small fluctuations.

In order to showcase the interaction of dilation with forgetting, we next systematically vary both (Fig.~\ref{fig:alpha3}a). For stochastic simulations, we explicitly draw random realizations of forgetting every step, whereas for exposure computations we add a decay term for integral exposure dynamics (see Appendices~\ref{app:simulation},\ref{app:exposure}). All of the stochastic trajectories have higher recall of nodes $R_\textsf{n}$ than edges $R_\textsf{e}$. Without forgetting ($\alpha=0$) the trajectory can get close to full recall for large enough dilation. In contrast, for high forgetting ($\alpha=0.01$) the walker can only remember the last $\sim$100 transitions, regardless of how long the walk was, and thus the learning trajectories look identical and terminate at fairly small recall. Forgetting thus severely limits the amount of memory available to the learner and the quality of mental models that they can form.

In order to compare directly the final recall of walkers at different dilation $D$ and forgetting $\alpha$, we perform a data collapse by combining the two variables into a single effective dilation:
\begin{align}
    D_\alpha=\frac{1}{\alpha \tau_{max}}\left( 1- e^{-\alpha \tau_{max} D} \right),
\end{align}
where $\tau_{max}$ is the number of sentences in a particular textbook. For small dilation $D$ the memory is dominated by learning so effective dilation tracks the actual dilation, whereas for large dilation $D$ the memory is dominated by forgetting and the total memory count plateaus at $\sim 1/\alpha$ (Fig.~\ref{fig:alpha3}b). This data collapse of effective dilation allows us to accurately predict node and edge recall across a wide range of actual dilation and forgetting (Fig.~\ref{fig:alpha3}c). We thus showed that recall is driven by the number of memories accumulated, which is limited by any amount of forgetting.

\section{The $\beta$ effect}
\subsection{Shuffling and precision}
\begin{figure}
	\includegraphics[width=\columnwidth]{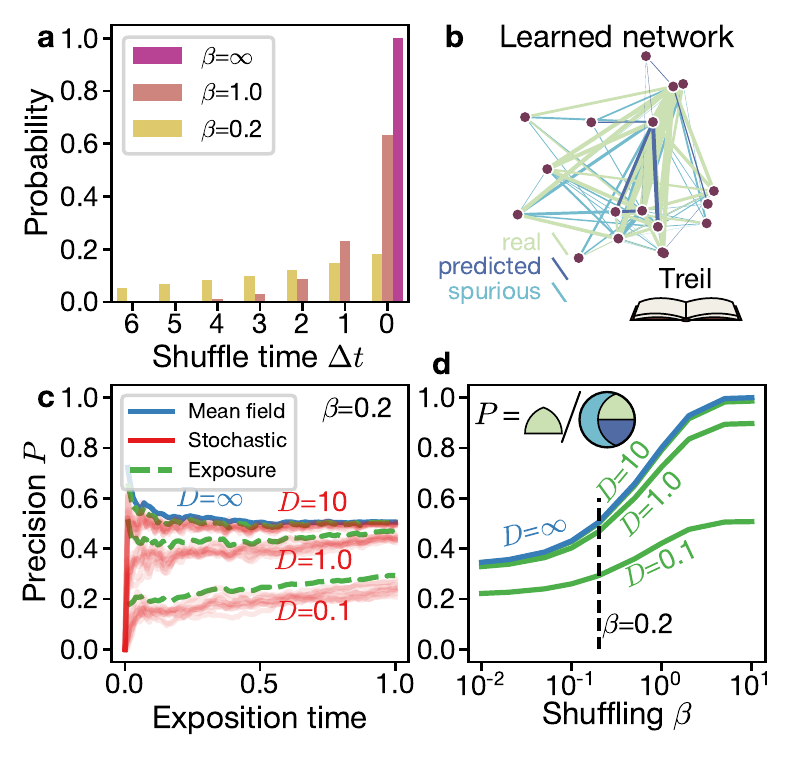}
	\caption{Memory shuffling leads to reduced mental model precision. (a) Each new memory connects the current node with one visited $\Delta t$ steps into the past, drawn randomly from a geometric distribution $p(\Delta t)\propto e^{-\beta \Delta t}$. (b) At an intermediate time the learned network consists of real edges (light green), predicted edges that would appear in the future (dark blue) and spurious edges (light blue). Only a small fragment of the network is shown for illustration. (c) Mental model precision $P$ throughout the exposition time at low shuffling $\beta$ and varying dilation $D$ from the mean field model (blue), 10 stochastic simulation replicas (red), and exposure theory (green). (d) Mental model precision across orders of magnitude of shuffling $\beta$ and varying $D$. The vertical line indicates the value $\beta=0.2$ used for panel (c).}
	\label{fig:beta1}
\end{figure}

The second effect we consider is the temporal shuffling of stimuli, which is characteristic of human memory processes. We follow the model established and experimentally validated in Ref.~\cite{lynn2020errors} and applied elsewhere \cite{lynn2020human, qian2021optimizing, Stiso2021}. We briefly recap it here. A human subject attempts to learn the network structure from the sequence of nodes $x(t)$ visited at each time step. In the absence of shuffling, each step adds a memory count to the entry $M_{ij}$ for $i=x(t)$, $j=x(t+1)$. However, remembering precisely the history of previously visited nodes requires significant mental resources; mental errors in recall are likely so that $i=x(t-\Delta t)$. The need to minimize errors $\Delta t$ is balanced in the brain with the need to minimize computational complexity \cite{ortega2013thermodynamics}. This trade-off can be expressed via the free energy principle, which predicts a geometric distribution of error sizes:
\begin{align}
	p(\Delta t)=(1-e^{-\beta})e^{-\beta\Delta t},
	\label{eqn:deltatdistr}
\end{align}
as illustrated in Fig.~\ref{fig:beta1}a for different values of the shuffling parameter $\beta\in\left[0,\infty\right)$. Ref.~\cite{lynn2020errors} proposed a way to measure the value of $\beta$ experimentally and found that for different human respondents it can be infinite (perfect memory), zero (full shuffling), or any finite value (partial shuffling). Therefore, in our computational model of the human learner, we need to consider a wide range of $\beta$, including the limiting behaviors $\beta\to 0$ and $\beta\to \infty$.

The $\beta$ model sets out the rules for the generation of erroneous memories from observing random walks on networks, but our investigation of mental errors is complicated in several ways with respect to previous results. First, we consider random walks on time-dependent networks. As shown in Fig.~\ref{fig:beta1}b, partway through learning a textbook network, some of the learned edges are real, whereas some are predictions of edges that would become real at a later point, and some are truly spurious. Second, we study the combination of mental errors with the under-sampling effect of finite-time random walks, as opposed to the infinite-time limit of Ref.~\cite{lynn2020errors}. Both of these complications are addressed in our simulations as described in Appendix~\ref{app:simulation} and exposure theory as described in Appendix~\ref{app:exposure}.

We first address the precision of learning real edges in the presence of shuffling (Fig.~\ref{fig:beta1}c). Throughout the exposition time, the precision remains nearly constant, but depends significantly on dilation $D$. As shown in Appendix~\ref{app:exposure}, the degradation of precision is mostly explained by failing to visit some of the nodes revealed by the textbook network, thus leading to the lack of \emph{any} mental model of transitions out of those nodes. At lower precision $D=0.1$ the under-sampling of nodes becomes particularly notable, leading to significant noise in the stochastic precision curve, and an over-estimation of the curve by exposure. Once the nodes have been visited, mostly by a dilation value of $D=10$, the precision follows the mean-field trajectory. Once a student is exposed to all of the concepts, after extensive study, they learn the network with the precision predicted by mean field theory.

How do increasing mental errors lead to loss of precision? At high $\beta\to \infty$ (near-perfect memory) and high dilation $D$ all nodes have been visited and only correct transitions are remembered; thus precision approaches 1 (Fig.~\ref{fig:beta1}d). At $\beta\to 0$ (full shuffling), the remembered edges randomly connect all remembered nodes; of all possible edges, many are real edges, and thus precision plateaus at some finite value $0<P<1$. Between the two extremes, the precision changes smoothly and monotonically; thus there is no ``optimal'' or ``threshold'' amount of memory shuffling. A learner with high shuffling would still learn all concepts, and all real connections---although along with all possible spurious connections.

\subsection{Edge prediction}
\begin{figure*}[ht]
	\includegraphics[width=\textwidth]{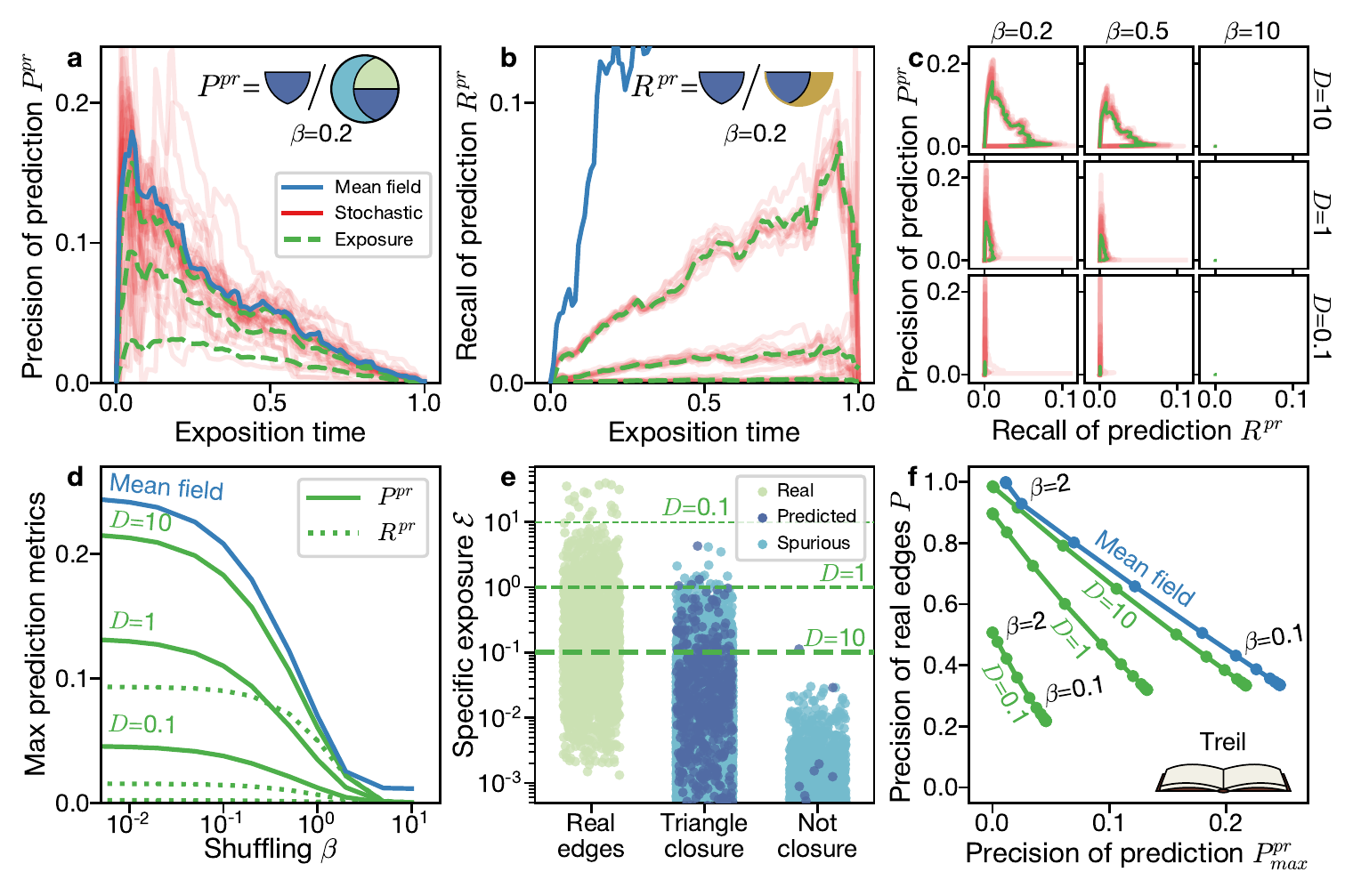}
	\caption{\textbf{Memory shuffling allows for limited prediction of future edges.} 
	(a) Trajectory of precision of prediction $P^{pr}$ with early peak and slow descent from mean field model (blue), 10 stochastic simulation replicas 
	(red), and exposure theory (green). (b) Trajectory of recall of precision 
	$R^{pr}$ with slow rise and rapid drop. The mean field curve (blue) is the
	limiting behavior at $D\to \infty$ but requires extremely high values of 
	$D$ to approach. (c) Prediction trajectories in $(R^{pr},P^{pr})$ space 
	across dilation $D$ and shuffling $\beta$. (d) Temporal maximum of 
	prediction precision (solid curves) and recall (dotted curves) at varying 
	shuffling $\beta$. The mean field temporal maximum of recall is 1 for any 
	finite $\beta$ (not shown). (e) Scatter plot of learned edges by specific 
	exposure at exposition time $\tau=0.6 \tau_{max}$ and shuffling 
	$\beta=0.2$. Colors indicate real, predicted, and spurious edges, while positions indicate real edges, triangle closures of real edges, and non-closures. Horizontal dashed lines indicate the boundary of learning at different values of the dilation parameter $D$. (f) Trade-off between the temporal maximum of precision of prediction $P^{pr}_{max}$ and the final precision of real edge learning $P$. The upper-left end of the curves corresponds to $\beta\to \infty$, whereas the lower-right end corresponds to $\beta\to 0$.}
	\label{fig:beta2}
\end{figure*}

In the next step of our investigation, we consider the prediction of future edges, which is inevitably a transient phenomenon. At the start of learning, there are no memories formed yet of either real or future edges, while by the end of learning, there is no learnable future. Therefore, all variation of prediction trajectories happens at intermediate times (Fig.~\ref{fig:beta2}). The trajectory shape is heavily modulated by the introduction of nodes over time (Fig.~\ref{fig:alpha1}a,b) since a shuffled random walk can only learn edges between the nodes that have already been introduced.

Similar to the learning of real edges, the learning of future edges (precision) can be characterized by either precision or recall (insets in Fig.~\ref{fig:beta2}a-b). The precision of prediction is the fraction of mental model probability weight that corresponds to future edges (as opposed to real and spurious edges). Compared to the recall of real nodes and edges (Fig.~\ref{fig:alpha1}a-b), the stochastic trajectories of precision of prediction show a much wider variation around the mean field and exposure curves. Unlike the recall of real edges, the precision of prediction is a fraction of two random numbers; thus the mean field curve does not serve as an upper bound, but merely an average trajectory. During exposition time, the precision of prediction has an early peak and a gradual fall-off (Fig.~\ref{fig:beta2}a). At early times, a fairly small fraction of nodes $N(t)$ has been introduced; thus the random walk memories are confined to relatively few $N(t)^2$ possible edges. Even among those nodes, many edges have not been introduced yet; thus a large fraction of probability weight falls on future edges, making them easy to discover and leading to the early peak. At later times, there are both more possible edges across which probability is spread and fewer future edges on which probability would be useful; together, these two factors result in a dwindling precision of prediction. 

The recall of prediction $R^{pr}$ is the fraction of all future edges learned and has the opposite trajectory shape: a gradual increase and a sharp drop-off (Fig.~\ref{fig:beta2}b). On one side, as more nodes and edges are introduced, they increase the range of edges that can be discovered by memory shuffling. On the other side, as the exposition time advances, more and more future edges become current edges, and the denominator of recall decreases, leading to the late peak. Compared to the precision of prediction $P^{pr}$, the stochastic trajectories follow exposure theory curves much more closely, but yet are very far below the mean field trajectory. In the mean field limit, any new edge between existing nodes can be predicted, but the likelihood of such a prediction at any finite dilation $D$ is extremely small.

Since the recall $R^{pr}$ and precision $P^{pr}$ of prediction follow opposite trends, they form a closed loop, starting and ending at zero and exhibiting a nearly linear dynamic trade-off in between (Fig.~\ref{fig:beta2}c). With advancement through the text, precision is traded for recall but at a steep rate (note the difference in scale on the two axes). As a learner starts reading the textbook, they first allocate a sizable part of their mental model to the prediction of future edges, but only end up covering a small fraction of them. 

How large can the precision and recall get throughout the book? Is there a value of shuffling $\beta$ that optimizes prediction? To answer these questions, we study the maximal prediction values and find that prediction gets monotonically higher with increasing shuffling or lower $\beta$ (Fig.~\ref{fig:beta2}d). For precision $P^{pr}$ the curve at $D=10$ closely follows the mean field curve since the under-sampling is mostly driven by the unvisited nodes. In contrast, for recall $R^{pr}$, the mean field curve is essentially always at 1, far above the plotting limits. Thus the dilation required to saturate the prediction recall seems markedly high. How does one reconcile the fact that the precision of prediction saturates by $D\sim 10$ with the fact that the recall does not?

The most common shuffling mistake in the model of human learning is confusing a random walk of length 2 for a random walk of length 1, i.e. $\Delta t=1$ or triangular closure. Therefore, at an intermediate point of the textbooks, we can separate all edges into three topological types: real edges, triangular closure of real edges, and all others (three point clouds in Fig.~\ref{fig:beta2}e, see Appendix~\ref{app:simulation} for definition). At the same time, the non-real edges are either predicted or spurious (dark and light blue in Fig.~\ref{fig:beta2}e). The triangular closure edges have notably higher specific exposure and can thus be reasonably discovered at $D\sim 1$, thus allocating a sizable fraction of mental model probability to future edges. In contrast, the non-closure edges only start being visible at $D\sim 10$. Since some of the future edges are not triangular closures, predicting them would require extremely high dilation $D>10^3$. This pattern of specific exposure of edges stratified by topological types explains why getting a substantial precision of prediction is easy, but getting a high recall is unlikely and should not be relied upon.

Lastly, how efficient is the trade-off between precision on real edges and prediction of future edges? The mental model probability is split between real, predicted, and spurious edges, but in what proportion? We find this trade-off to be essentially linear but limited (Fig.~\ref{fig:beta2}f). At high values of $\beta$ learning is precise and lies in the top-left corner. At low values of $\beta$ learning is fully shuffled, and both metrics approach finite values that are dependent on the density of real and future edges. For the Treil textbook, as shown here, the trade-off between $P$ and $P^{pr}_{max}$ is about $3:1$, but gets even steeper at lower dilation (other textbooks have a similar pattern; see Fig.~\ref{fig:tradeoff}). The memory shuffling thus affords learners a limited ability to predict future edges at the cost of precision of real edges.

\section{The $\gamma$ effect}
\subsection{Reinforcement and slowdown}
\begin{figure*}
	\includegraphics[width=\textwidth]{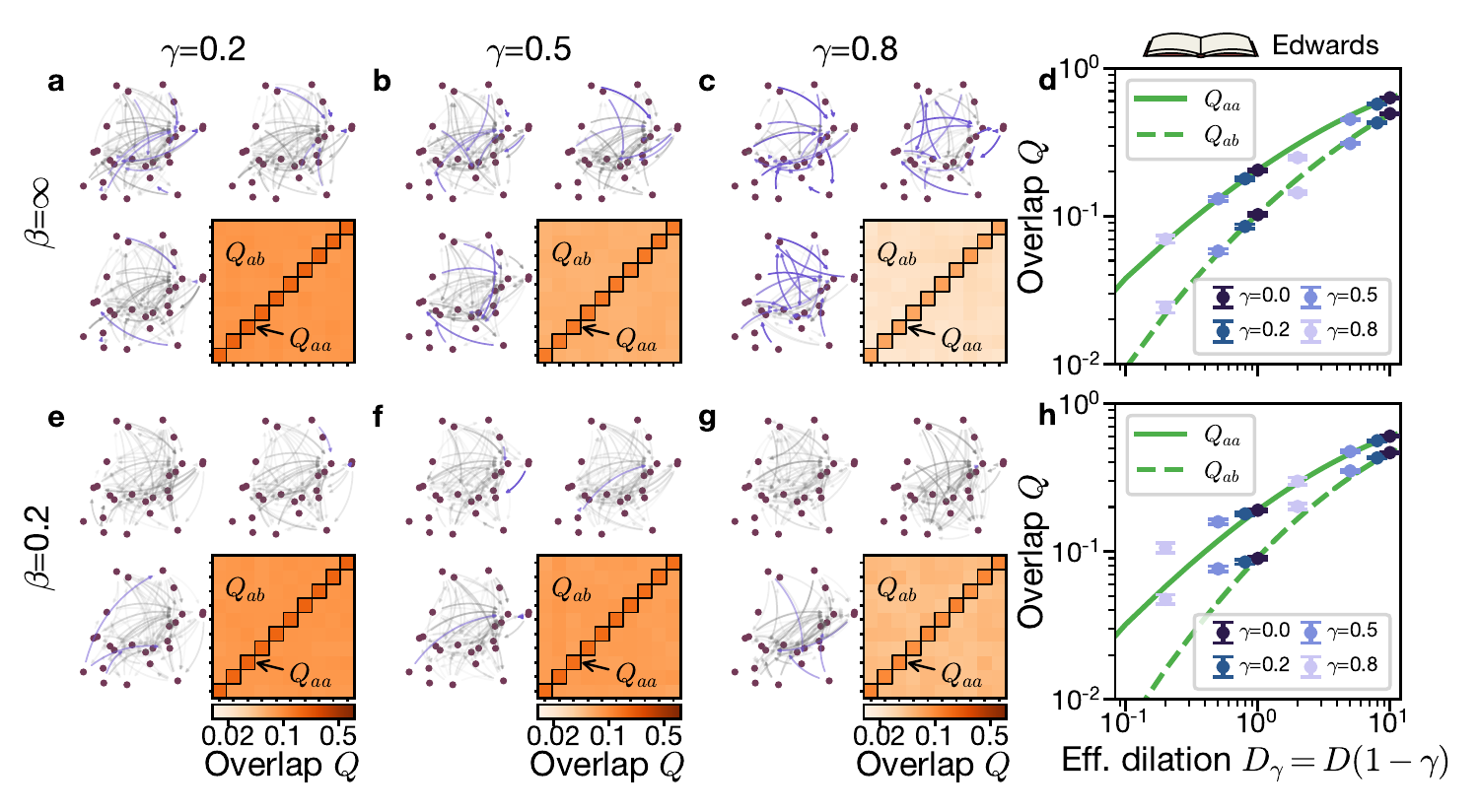}
	\caption{\textbf{Reinforcement of random walks causes a slowdown of exploration, but shuffling mitigates it.} (abc, efg) Illustration of learned transition networks at varying reinforcement $\gamma$ (columns) and shuffling $\beta$ (rows), but identical dilation $D=1.0$. Arrow transparency indicates the transition probability $\hat{T}_{ij}$, with most likely edges $\hat{T}_{ij}>0.4$ shown in purple. The heatmap shows the overlap matrix $\mathbf{Q}$ of 10 replicas at 	given values of the $\gamma$ and $\beta$ parameters. Matrix diagonal corresponds to self-overlap $Q_{aa}$, off-diagonal corresponds to cross-overlap between the replicas $Q_{ab}$. (d,h) Overlap metrics plotted against \emph{effective} dilation $D_\gamma=D(1-\gamma)$, for simulations performed only at two values of dilation $D=10^0,10^1$. Self-overlap is shown by the upper marker in each pair for simulations and the solid curve for exposure prediction. Cross-overlap is shown by the lower marker in each pair for simulations and the dashed curve for the exposure prediction.}
	\label{fig:gamma1}
\end{figure*}

The third effect we consider is the random walk reinforcement, which makes already existing memories stronger. While the $\alpha$ and $\beta$ effects only modify the way memories behave, the $\gamma$ effect changes the actual random walk steps. The $\gamma$ effect accounts for the tendency to revisit the same edges that one remembers, inspired by models in Refs.~\cite{iacopini2018network, lydon2021hunters}. While in those studies the underlying weights of the transition matrices were modified, here we instead fix the probability of taking a step on the textbook network or the mental model. Mathematically, if the learner is on a node $i$ with pre-existing memories, they choose their next step with probability:
\begin{align}
	P(j|i)=(1-\gamma)T_{ij}(\tau)+\gamma \hat{T}_{ij},
	\label{eqn:reinforcement}
\end{align}
where $T_{ij}$ is the transition matrix of the textbook, $\hat{T}_{ij}$ is the mental model of transitions, and $\gamma\in[0,1]$ is the mixing factor. At $\gamma=0$ the random walk follows the textbook network, whereas at $\gamma=1$ the random walk exclusively retraces the existing edges if any are remembered. Hence, the $\gamma$ parameter regulates the degree of positive feedback, since following known edges creates more memories of those edges, and makes them more likely to be traversed again.

How exactly does reinforcement affect network exploration and mental model building? Are the mental models built by different learners consistent with each other? It is important to recognize the space of possible mental models as high dimensional. While in a one-dimensional dynamical system with positive feedback the state variable just grows, in the high-dimensional space of possible mental models $\hat{T}_{ij}$, the early random walk steps select random edges from the textbook network, and the later steps reinforce the memories of those edges. The combination of early random selection with positive feedback results in many different transition networks $\hat{T}_{ij}$ formed in independent stochastic replica simulations at varying levels of reinforcement $\gamma$ (Fig.~\ref{fig:gamma1}a-c). As $\gamma$ gets higher, the resulting networks get less similar to each other and get more high-probability edges (purple arrows for $\hat{T}_{ij}>0.4$). In order to compare the independent replica runs, we use an overlap metric inspired by the spin glass literature 
\cite{castellani2005spin}:
\begin{align}
	Q_{ab}=\frac{1}{m}\sum\limits_{ij}[T_{ij}>0][\hat{T}^a_{ij}>0]
	[\hat{T}^b_{ij}>0],
	\label{eqn:overlap}
\end{align} 
where $m$ is the number of directed edges in the textbook network and the 
superscripts $a,b$ correspond to the replica indices. When the indices are the 
same, the self-overlap $Q_{aa}$ measures how many edges are shared between the 
taught network and the learned mental model. When the indices are different, 
the cross-overlap $Q_{ab}$ accounts for the overlap between two replicas. The self- and cross-overlap remain consistent across replicas but different from each other (heatmaps in Fig.~\ref{fig:gamma1}a-c). As $\gamma$ increases, both overlap metrics decrease (more pale color on the heatmap): that is, reinforced random walks explore less of the taught network, and build mental models less similar to each other.

How significant is the drop in exploration and overlap caused by a growing value of reinforcement $\gamma$? The only way for the learner to discover new edges is to take steps along the textbook network, which happens with probability $(1-\gamma)$ (note that some steps along the textbook network still retrace older memories). We thus hypothesize that exploration statistics at dilation $D$ and reinforcement $\gamma$, as measured by overlap $Q_{ab}$, would follow the un-reinforced statistics at lower \emph{effective} dilation $D_\gamma=D(1-\gamma)$. For the un-reinforced statistics, exposure theory predicts both overlap curves shown in Fig.~\ref{fig:gamma1}d. In order to test the effective dilation hypothesis, we perform reinforced random walk simulations at varying $\gamma$ but only two values of dilation $D=10^0,10^1$ and compute the self- and cross-overlap statistics. By converting each pair of parameters into a single parameter $D,\gamma \to D_\gamma$, the overlap data collapse onto the curve predicted by exposure theory (Fig.~\ref{fig:gamma1}d). We thus show that reinforced random walks slow down exploration of the networks and make less consistent mental models, both in proportion to reinforcement.

The slowdown effect relies on reinforcement of the revisited edges, which requires remembering them correctly. However, it is possible that the learner would follow their own memories, but not form memories correctly: that is, the $\gamma$ effect can coexist with the $\beta$ effect. In order to test whether the $\beta$ effect breaks the feedback loop, we perform simulations that simultaneously take into account reinforcement and memory shuffling. The mental models appear more consistent with each other, with fewer strong (purple) edges emerging (Fig.~\ref{fig:gamma1}e-g). The overlap heatmaps still get paler with growing $\gamma$, but the effect is much less pronounced. For simulations at $D=1.0$ and $\gamma>0$ the overlap metrics are lower than at $\gamma=0$ but higher than predicted by effective dilation (Fig.~\ref{fig:gamma1}h). Memory shuffling thus can partially mitigate the reinforcement slowdown by breaking the positive feedback loop. A student with these two mental effects would thus be able to discover additional edges compared to exploration with reinforcement alone.

\subsection{Symmetry breaking}
\begin{figure*}
	\includegraphics[width=\textwidth]{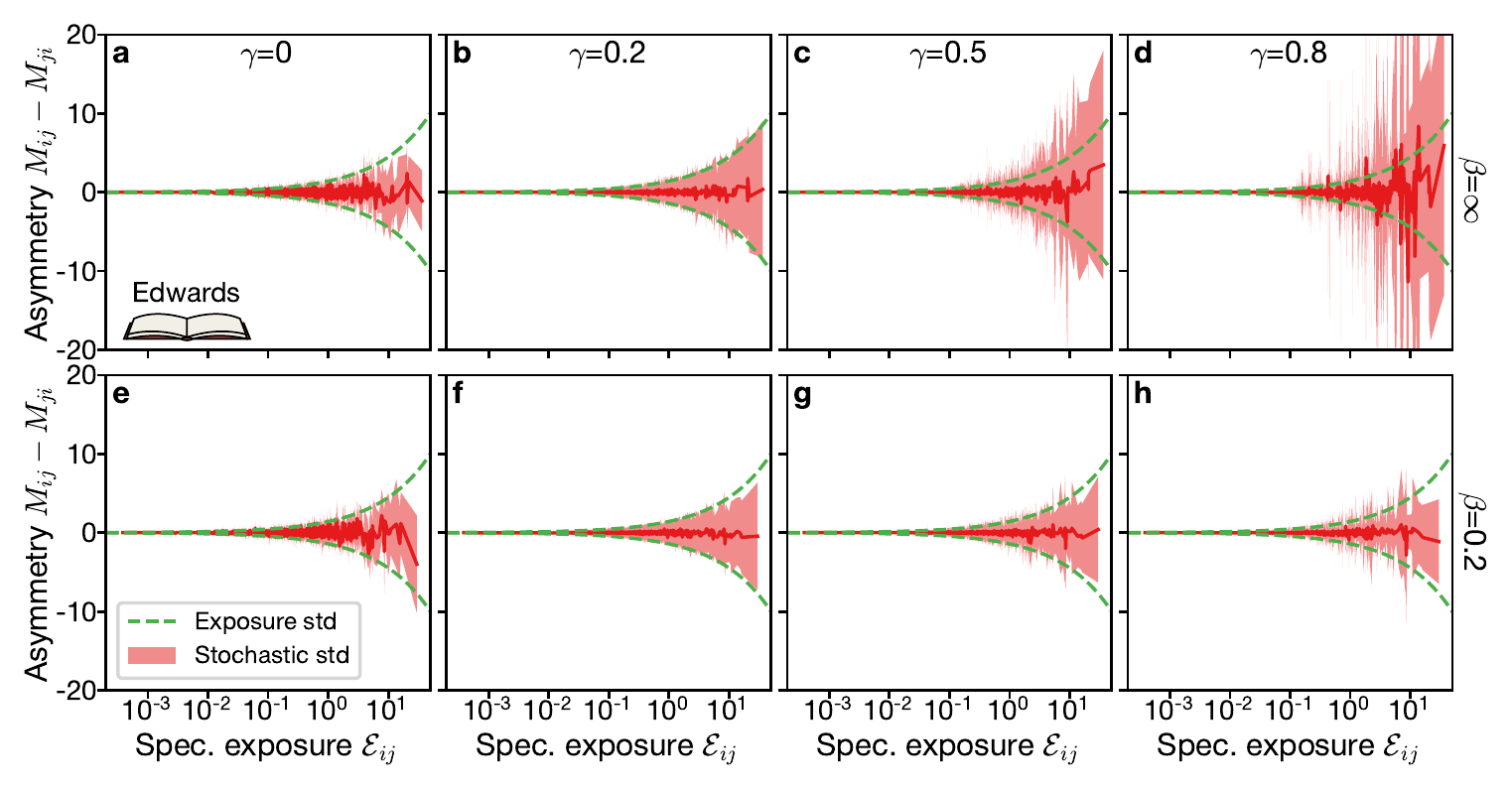}
	\caption{\textbf{Reinforcement causes edge reciprocity breaking beyond the exposure 
	null model.} Columns correspond to varying reinforcement $\gamma$, rows to 
	shuffling $\beta$. The null model mean difference is always zero, while 
	green dashed curves show the standard deviation predicted by exposure. The 
	red curve shows the stochastic mean, while the red shading shows the 
	stochastic standard deviation over 100 replicas. All simulations and 
	analytical computations are performed at dilation $D=1.0$.}
	\label{fig:gamma2}
\end{figure*}

We previously noted that reinforcement can produce very strong edges in the mental model (Fig.~\ref{fig:gamma1}c). At the same time, these edges are not reciprocal: while the underlying textbook network is undirected and thus is described by symmetric matrices, the learned mental model appears to be breaking symmetry significantly. In order to quantify the degree of symmetry breaking, we must first establish the null expectation. Exposure theory predicts the number of memories to follow the Poisson distribution: $M_{ij}=\textsf{Pois}(D\mathcal{E}_{ij})$, where the exposure matrix is symmetric. The memory counts in an edge $M_{ij}$ and its reciprocal $M_{ji}$ thus follow identical distributions, but are independent from each other. The asymmetry $M_{ij}-M_{ji}$ is then a random number with a mean of zero and a variance twice that of each edge, i.e. $2 D\mathcal{E}_{ij}$. As specific exposure of edges spans nearly five orders of magnitude across the set of textbooks, the expected asymmetry for edges also varies significantly.

Increasing the reinforcement $\gamma$ significantly changes the statistics of asymmetry. Without reinforcement at $\gamma=0$, asymmetry statistics fall within the standard deviation envelope $\pm\sqrt{2 D\mathcal{E}_{ij}}$ predicted by exposure theory (Fig.~\ref{fig:gamma2}a). As reinforcement $\gamma$ increases, the asymmetry leaves the envelope both above and below zero so that the symmetry of the transition matrix can be significantly broken in either direction (Fig.~\ref{fig:gamma2}b-d). But what if the random walk steps are not remembered correctly, for instance if the $\beta$ effect (shuffling) is also present? In this case, the asymmetry effect reduces significantly and the statistics mostly fit within the predicted envelope at all values of $\gamma$ (Fig.~\ref{fig:gamma2}e-h). Memory shuffling is thus able to mitigate the positive feedback loop caused by reinforcement. A student with shuffled memories thus does not falsely infer strong directionality of conceptual connections that is absent in the textbook network.

\section{Discussion}
In this paper we set out to describe the translation of taught semantic networks into learned ones, sculpted by the dual forces of finite learner effort and the specific effects of human memory. Our conclusions about the role of the three memory effects can be readily related to other studies, as shown below. At the same time, converting teaching into learning is the primary goal of education, for which our study provides a missing link. Lastly, this work substantially expands upon exposure theory and allows us to map out the technical limitations and further open avenues.

\subsection{Memory effects}
Alongside under-sampling, we also account for several memory effects reported in the literature. The first effect investigated is the $\alpha$ effect, which corresponds to a uniform forgetting rate of all formed memories. While the forgetting process is not biased towards some nodes and edges over others, the less connected nodes and weaker edges would naturally have fewer memory counts accrued and thus are more likely to be forgotten completely. The exact time course of forgetting memories has been a longstanding subject of debate, with several empirical functional forms---such as exponential, power law, or hyperbolic---having nearly equal data support \cite{wozniak1995two, rubin1996one, rubin1999precise}. A uniform forgetting rate corresponds to an exponential forgetting curve, which is thus plausible. This forgetting model leads to the steady-state memory size of about $1/\alpha$, which has several implications.

On one side, forgetting limits the memory sample size from which a learner constructs their mental model, and thus puts a strict limit on node and edge recall. The forgetting rate is connected to other cognitive processes such as event segmentation and is different across human subjects \cite{jafarpour2022event}; thus different learners are expected to hold different amount of memories. On the other side, the memories are restricted to the most recent random walk steps, which in the case of growing textbook networks corresponds to sampling essentially the full network available by the end of the book. The forgetting effect thus appears detrimental if remembering all nodes and edges forever is considered to be a goal and a virtue. However, neurophysiological evidence into the mechanisms of forgetting suggests that moderate forgetting is necessary and expected across many organisms \cite{ryan2022forgetting}. Forgetting can have epistemological benefits as it prunes memory of low-importance edges or of edges that do not exist anymore, leading to a clearer and simpler mental model \cite{michaelian2011epistemology}.

The second effect investigated is the $\beta$ effect, which represents mental errors in the form of shuffling. The $\beta$ effect does not change the \emph{number} of memories formed, but does affect their placement. In prior work, shuffling was found to enhance the relative weight of edges within network communities and decrease the relative weight of edges between communities \cite{lynn2020errors}. This effect is driven by the differential \emph{leakage} of probability from the real within- and between-community edges into the spurious edges, causing the mental model to have a finite precision. Networks with pronounced community structure have lower leakage, and minimizing leakage is hypothesized to be a selection pressure on the architecture of communication networks \cite{lynn2020human}. Carefully re-weighing the edges of the input network to emphasize communities and de-emphasize connections between them can to some degree mitigate the leakage effect and lead to more precise learning \cite{qian2021optimizing}.

The combination of memory shuffling and the temporal nature of the textbook networks can lead to learning edges before they are introduced, and thus in effect ``predicting'' them. Edge prediction is a common problem in network theory: usually a fraction of edges is used to train a computer algorithm that attempts to predict the other, ``holdout'' edges \cite{lu2011survey, ghasemian2020stacking}. Some of the algorithms themselves rely on a random walk as a local or quasi-local process to score the possible missing edges \cite{liu2010link, lu2011survey, berahmand2021preference}. The prediction based on shuffling is qualitatively different: throughout the random walk, humans automatically and randomly add memories of edges that were not directly traversed, and some of those edges appear in the textbook later (like those in the ``holdout'' set). Here it is important to clarify that the ``prediction'' of edges does not follow a scoring algorithm, but is rather guessing. The efficiency of such a prediction depends on the authorial choices of the order in which concepts and connections are introduced in a particular book---some textbook might be more predictable than others. Precision and recall of prediction form a dynamical trade-off throughout the exposition, wherein the precision of predictions gives way to their amount (recall). The temporal peak of prediction forms yet another trade-off with the precision of real edges, but the conversion ratio is quite steep. While edge prediction is a robust effect, and can to some degree be engineered through text ordering, it is only a secondary benefit that partially compensates for loss of precision.

The third effect investigated is the $\gamma$ effect, or the reinforcement of the random walk by its own memory, which has been examined before. An edge reinforcement model was previously used to explain the strongly sublinear statistics and correlations in the discovery of novelties on a network \cite{iacopini2018network}. A similar model previously explained the tendency of some curiosity-driven Wikipedia readers to return to already known concepts and close the remaining knowledge gaps instead of exploring new areas \cite{lydon2021hunters}. In our case, adding any amount of reinforcement not only slows down network exploration proportionally, but also leads to spontaneous reciprocity symmetry breaking: the learners infer edge directionality even if the textbooks did not have any. At the same time, adding memory shuffling (as operationalized in the $\beta$ effect) allows the learner to still form new memories. Under the combination of these two effects, the learner can revisit known parts of the network but still discover new connections there.

The mental model built under reinforcement is not determined purely by the textbook network; instead, it builds upon the random choice of the initial few steps and is thus path dependent \cite{page2006path}. The idea of path dependence first became prominent in explaining the positive feedback in economic systems \cite{arthur1994increasing} and since has been fruitfully applied across other social sciences \cite{magnusson2009evolution} and especially in the study of persistent institutions \cite{bednar2015choosing}. In mathematical modeling of path dependence, mere enumeration of possible system states presents a significant problem. At the same time, in network models the set of nodes defines explicitly the set of possible edges that can be learned, making path dependence high-dimensional but still tractable. Within our model we showed that increasing reinforcement leads to learners forming less similar mental models as measured by overlap. Going forward, we envision network models to be an especially useful platform to study path dependent phenomena more broadly, driven by a variety of other mental effects.

Do the memory effects represent ``failures'' of human learning as compared to automatic computer learners performing optimal inference? Across the three memory effects, we find that deviating from the ``ideal case'' ($\alpha=0, \beta=\infty, \gamma=0$) degrades the quality of learning as measured by recall, precision, and overlap, yet most humans deviate from that cognitive regime. Some normal amount of forgetting is argued to be beneficial for decluttering our mental models and avoiding overfitting to noise \cite{michaelian2011epistemology}. Persistent shuffling of stimuli should degrade the human learning of networks, but instead it serves as a selection pressure on the structure of cognitive networks that humans build in the first place \cite{lynn2020human}. The tendency to revisit known edges in curiosity-driven exploration is not a limitation of learning but a mere facet of the many styles of curiosity \cite{lydon2021hunters}. Within these three characterized effects, and possibly along other uncharted axes, humans show natural variability. As instructors we do not get to choose the memory parameters of our students; at most we get a rough measure of what those parameters are so that we can adjust the teaching structure accordingly.

Another contribution of exposure theory to the learning sciences is the bridging of implicit learning mechanisms to the development of network structures that represent the learner's explicit knowledge of a domain. Implicit learning typically refers to the acquisition of knowledge that occurs through passive exposure to information in the environment. Humans are associative learners---we pick up statistical information of relationships through mere exposure \cite{Saffran_Aslin_Newport_1996} and it has been shown that there are stable individual differences in implicit learning ability \cite{kalra_evidence_2019}. In the educational context, implicit learning could occur when a learner is exposed to the patterns of their learning environment, for instance, when passively reading (or skimming) a textbook.

Given that humans readily pick up the associations in their environment, an important question is: How do learners use these associative patterns to build up sophisticated, large-scale knowledge structures? We suggest that exposure theory can provide potential answers to this question. Memory research has long-established that human memory does not behave like a computer that stores replicas of one's perceptual experiences \cite{pan_acquiring_2020}. In particular, memory effects serve an adaptive function in crafting the specific memory structures of humans in a way that optimizes later retrieval and learning. Forgetting and shuffling serve key functions in retaining core structural features of the taught network at the expense of precision. These two variables in our model are akin to empirically measured forgetting processes and reactivation of memories in random sequences (i.e., shuffling that occurs naturally in spontaneous thought), which respectively enable gist-extraction and abstraction of knowledge in human learning \cite{storm_learners_2016, lynn2020human, zeng2021tracking} as well as the optimization of later retrieval of competing concepts \cite{hulbert_neural_2015}. Reinforcement provides an explanation for how the prior history of a learner's trajectory affects later learning and retrieval, which in turn shapes the memory structure. This explanation aligns with previous investigations which show that retrieval strengthens the storage and retrieval strength of previously learned material \cite{bjork_new_1992, roediger_test-enhanced_2006}. Broadly, exposure theory as modified by memory effects provides one framework for understanding implicit learning: how passive exposure to the structure of textbooks or other taught materials are filtered through key features of human memory and develop into more explicit forms of knowledge structures.

\subsection{Educational implications}
It is widely understood that different teaching methodologies of the same material can result in different learning across matched student cohorts \cite{deslauriers2011improved, denervaud2021education}. For the same teaching, in any given classroom the students are going to put a different amount of effort towards learning \cite{palazzo2010patterns}, and are remarkably unresponsive to interventions to increase that effort \cite{oreopoulos2019remarkable}. At the same time, humans naturally vary in their memory and in their curiosity along multiple axes \cite{lynn2020errors, lydon2021hunters, nilsson2021structural, jafarpour2022event}. Under these wide-ranging conditions, how does the taught material become the learned material? How can the instructor adjust the material or presentation to increase learning?

A range of previous studies have considered the semantic networks of concepts held by the teachers or textbooks \cite{yun2018extraction, christianson2020architecture, vukic2020structural}, while others examined networks constructed by students \cite{corbett2010cognitive, siew2019using, koponen2018genealogy, lommi2019landmarks}. Yet rarely have the two been analyzed jointly. Our modeling results here suggest that the same taught network can result in many different learned networks, and further that the mapping can be described in terms of just a few parameters of learners' effort and memory. Validating these predictions would require a carefully controlled experimental study where both teacher and student semantic networks are assessed simultaneously.

A core idea in the memory and science-of-learning literature is that \emph{learning} and \emph{performance} are distinct constructs. Their differences can make it challenging to distinguish between information that was better learned and information that is easy to retrieve. Such a distinction is made in the so-called \emph{new theory of disuse} that discusses the differences between storage strength (learning) and retrieval strength (performance) of items in memory \cite{bjork_new_1992}. While our approach does not explain differences in the learners' ability to retrieve information from their mental models, it does provide a much-needed formalism for representing the learned memory structure that is obtained from the taught structure---with a careful consideration of common memory errors committed by humans. Establishing the baseline learned structure provides an important foundation for further modeling of retrieval or recall processes \cite{polyn_context_2009, hills_optimal_2012, abbott_random_2015, naim2020fundamental} that will have implications for explaining performance differences in educational settings. 

The framework of exposure theory can be useful not only in experimental validation of learning predictions, but also in the design of teaching materials. Exposure theory provides a detailed map of heterogeneous concepts and connections, which can serve as an early feedback mechanism during the development of a textbook or syllabus. Are the concepts deemed important by the author actually properly emphasized in the text? What is the picture that the most cursory student would get out of the course? When we consider individual differences in learners' dispositions and cognitive abilities, what is the ``range'' of their learned networks? In other words, what is the variance in the network structures acquired from the learning materials, and what aspects of their learned networks are consistent across learners? These questions highlight a potential use-case of our approach: textbook design could be aimed towards building learning landscapes that either reduce variance across the network characteristics of memory structures attained across learners of different dispositions, or ensure that a large proportion of learners are likely to acquire the core knowledge of a given discipline (as defined by the author(s)). While we do not advocate using exposure metrics as a sole method of ``optimizing'' the teaching materials, our study opens the conversation about the materials design.

Finally, aside from enhancing teaching material design, it would also be possible to use exposure metrics to enhance assessment design, or at least tailor the assessment in a way that aligns with the textbook structure and the population of learned network structures across students. Could an assessment be evaluated based on whether its sampling of to-be-tested concepts or associations across the network can effectively distinguish between students who perform well versus poorly in the class?

\subsection{Methodological considerations}
In this paper we extend the exposure theory formalism to account for three orthogonal and empirically motivated memory effects. We also expect exposure theory to be able to handle other effects so long as network learning remains ergodic; that is, so long as the edge visitation probabilities quickly relax to the values dictated by the textbook network \cite{klishin2022exposure}. The $\alpha$ and $\beta$ effects modify memory probabilities but do not break erdogicity, thus retaining accuracy. The $\gamma$ effect explicitly breaks ergodicity by adding path dependence \cite{page2006path}, which limits our ability to make predictions. Visitation of new edges is still driven by the textbook network, which allows us to make accurate predictions of overlap. In contrast, reinforcement of existing edges and reciprocity symmetry breaking are path dependent, so exposure theory specifies the null model of asymmetry that is violated, but not the exact nature of that violation.

While tracking exposure was originally conceived of as a cheap and accurate numerical proxy for stochastic simulations, the exposure value can have other applications. The mapping from the exposure value to the visit probability is a nonlinear convex function: here it is exponential (Eqn.~\ref{eqn:probvisit}), but theoretically it can take other functional forms. For example, Ref.~\cite{kollepara2021unmasking} connects the exposure to viral loads with the probability of developing an infectious disease such as COVID-19 and uses curve convexity to argue for a super-linear benefit of mask-wearing to prevent infections. Studies of the spread of social behaviors center on the mechanism of complex contagion, in which an individual needs to be exposed to a behavior multiple times from different sources in order to adopt it themselves \cite{romero2011differences, guilbeault2018complex}. Similar to those studies, the functional form of the exposure-to-probability map underlies the global features of the spreading dynamics, including the Jensen bound and the trade-off between exploration speed and prioritization.

The results of this paper rely on the modeling choices of converting textbooks into networks and using random walks to explore those networks. In order to map out the substrate for random walks, we convert the network measurements of textbooks in Ref.~\cite{christianson2020architecture} into dynamic networks with a simple rule: each network edge appears immediately at full strength $A_{ij}$ as soon as the exposition time $\tau$ reaches the filtration order value $F_{ij}$. However, the semantic connection between two concepts might be limited to only one chapter of the book, in which case the edge between them should only exist for a finite time. Statistics of random walks and other spreading processes are known to change significantly when the timescales of the random walk step and network evolution are matched \cite{perra2012random} or when network the evolution is intermittent \cite{allen2022compression}. Further, the introduction of new concept connections in the text often guides the learner to explore them, thus biasing the random walk towards the freshly-introduced parts of the network \cite{snyder2008form, mather2013novelty}. Lastly, simple random walks are known to be fairly inefficient means of exploring random networks, and many more sophisticated algorithms are available \cite{asztalos2010network, bonaventura2014characteristic, dearruda2017knowledge}. While the combination of these limitations suggests that random walks are at best an incomplete model of a learning process, exposure theory greatly speeds up the analysis of this model in bypassing costly stochastic simulations. We therefore pose this study as an important baseline against which to compare the effect of additional learning mechanisms.

\section{Conclusions}
In this paper we propose a model of how taught semantic networks turn into learned ones, usually non-exactly. We consider two main limitations of learning: the under-sampling effect due to learning for a finite time and three types of memory imperfections individually validated in the literature. We expand the domain of exposure theory to accurately predict the interplay of under-sampling with diverse memory effects at a fraction of the computational cost required by stochastic simulations. While prior work mapped out separately the semantic networks of teachers and learners, our findings suggest the possible shapes of network distortions in the learning process that can be investigated experimentally. Exposure-based analysis can be used to predict the chance of learning concepts and connections from instructional materials, and thus can be used as a design tool for those materials.

\section*{Acknowledgments}
The authors would like to thank C.W.~Lynn and X.~Xia for discussions about the modeling, as well as L.~Dourte, N.~Finkelstein, D.~Pritchard for discussions on the education research literature. The computational workflow in general and data management in particular for this work was primarily supported by the Signac data management framework \cite{signac_commat,signac_zenodo}. This research was funded by the Army Research Office (DCIST-W911NF-17-2-0181) and the National Institute of Health (R21-MH-106799). The content is solely the responsibility of the authors and does not
necessarily represent the official views of any of the funding agencies.

\section*{Citation diversity statement}
%numbers computed on June 30, 2022

Recent work in several fields of science has identified a bias in citation practices such that papers from women and other minority scholars are under-cited relative to the number of such papers in the field \cite{mitchell2013gendered,dion2018gendered,caplar2017quantitative, 	maliniak2013gender, dworkin2020extent, bertolero2021racial, wang2021gendered, chatterjee2021gender, fulvio2021imbalance, teich2021citation}. Here we sought to proactively consider choosing references that reflect the diversity of the field in thought, form of contribution, gender, race, ethnicity, and other factors. First, we obtained the predicted gender of the first and last author of each reference by using databases that store the probability of a first name being carried by a woman \cite{dworkin2020extent,zhou_dale_2020_3672110}. By this measure (excluding references in this paragraph and self-citations to the first and last authors of our current paper), our references contain 20.41\% woman(first)/woman(last), 13.24\% man/woman, 12.41\% woman/man, and 53.94\% man/man. This method is limited in that a) names, pronouns, and social media profiles used to construct the databases may not, in every case, be indicative of gender identity and b) it cannot account for intersex, non-binary, or transgender people. Second, we obtained predicted racial/ethnic category of the first and last author of each reference by databases that store the probability of a first and last name being carried by an author of color \cite{ambekar2009name, sood2018predicting}. By this measure (and excluding self-citations), our references contain 11.15\% author of color (first)/author of color(last), 10.77\% white author/author of color, 17.22\% author of color/white author, and 60.85\% white author/white author. This method is limited in that a) names and Florida Voter Data to make the predictions may not be indicative of racial/ethnic identity, and b) it cannot account for Indigenous and mixed-race authors, or those who may face differential biases due to the ambiguous racialization or ethnicization of their names.  We look forward to future work that could help us to better understand how to support equitable practices in science.

\appendix

\section{Textbook networks statistics}
\label{app:textbooks}
\begin{table}[]
    \centering
    \begin{tabular}{lcccc}
    \toprule
      Book &  Nodes $n$ &  Edges $m$ &  $\tau_{max}$ &  $t_{corr}$ \\
    \midrule
         Treil &    278 &   7106 &     6681 &  2.66 \\
         Axler &    217 &   8458 &     4220 &  1.78 \\
       Edwards &    146 &   4322 &     2066 &  2.06 \\
          Lang &    179 &   5174 &     3958 &  1.94 \\
      Petersen &    244 &   8940 &     5742 &  1.83 \\
      Robbiano &    219 &   8086 &     2944 &  1.92 \\
     Bretscher &    384 &  11914 &    12703 &  3.12 \\
         Greub &    275 &   7108 &     6841 &  3.34 \\
     Hefferson &    399 &  13042 &     8046 &  3.60 \\
        Strang &    453 &  15512 &    10965 &  2.70 \\
    \bottomrule
    \end{tabular} 
    \caption{\textbf{Basic statistics of the textbooks used in the study.} $\tau_{max}$ is the number of sentences; $t_{corr}$ is the correlation time on the full network.}
    \label{tab:bookstats}
\end{table}

In this paper we consider 10 popular linear algebra textbooks (Table~\ref{tab:bookstats}). Each textbook was written by a single author, whose last name we use as a shorthand for the book throughout the paper. The network extraction procedure is described in Ref.~\cite{christianson2020architecture}. Each textbook network consists of $n$ nodes and $m$ directed reciprocal edges: we count edges $i\to j$ and $j\to i$ separately since they can be learned separately. The length of each textbook is measured by the number of sentences $\tau_{max}$ so that the number of random steps a learner with dilation $D$ would take is $t_{max}=D\tau_{max}$. The random walk correlation time $t_{corr}$ is determined from the second eigenvalue of the transition matrix as described in Ref.~\cite{klishin2022exposure}. The assumptions of exposure theory are fulfilled so long as $t_{corr}\ll t_{max}$, which is the case for all textbooks in the range of dilation we consider in this paper.

\section{Stochastic simulations}
\label{app:simulation}
\subsection{Baseline simulations}
The stochastic simulations are implemented via a custom code written in Python. We first describe the baseline simulation, and then the necessary algorithmic modifications to account for the three memory effects.

We initialize the memory matrix $\mathbf{M}$ as a sparse, integer-valued $N\times N$ matrix with no entries. Since the indices $i,j$ increment in the order of appearance, we start the random walks at the node $i=0$ so that that node is guaranteed to have edges at the early stages of network growth. At each time step $t$ we compute the evolution time $\tau=t/D$. Since the random walker is known to be at node $i$, we only need to evaluate one row of the transition matrix $P(j|i)$ following Eqn.~\ref{eqn:Pji}.

The most computationally expensive step in the random walk algorithm is the generation of pseudorandom numbers. At the same time, for each realization of a random walk, we might need to compute different time-dependent metrics based on the memory matrix $\mathbf{M}$. It is not efficient to store too many snapshots of $\mathbf{M}$ at different time points and stochastic realizations on the hard drive, so we instead store the random walk trajectory and reconstruct it on demand. We denote $x(t)$ to be the node $i$ at which the random walker is located at time $t$, and $x'(t)$ to be the node from which the walker \emph{remembers} to have arrived from. Without any memory effects, the update procedure is as follows:
\begin{align}
	\left.P(j|i)\right|_{i=x(t)}\to x(t+1);\quad x'(t+1)=x(t),
\end{align}
where the operator $\to$ denotes drawing a pseudorandom realization from the probability distribution. Once the trajectories have been computed, the memory matrix can be reconstructed as follows:
\begin{align}
	M_{ij}(t)=\sum\limits_{t'=1}^{t}[i=x'(t')][j=x(t')].
	\label{eqn:recon}
\end{align}
The computational benchmark of a direct simulation versus a reconstruction is presented in the Supplementary Materials of Ref.~\cite{klishin2022exposure}, but typically results in a reduction of computation time by a factor of $10^1..10^2$.

In order to accumulate statistics that support our main results, we perform several thousand simulations at different parameter values (including the pseudorandom seed), forming several series of computational experiments. We organize the computational workflow in general and data management in particular with the Signac data management framework \cite{signac_commat, signac_zenodo}.

\subsection{The $\alpha$ effect}
In presence of the $\alpha$ effect every memory is forgotten at every step with a uniform probability $\alpha$. Given the memory count $M_{ij}$ in a given cell, the number of memories forgotten is a binomial random number $B(M_{ij},\alpha)$. If some cell already has zero memories, then none can be forgotten. We therefore only draw the pseudorandom realizations for cells $M_{ij}$ with non-zero entries. Since the whole memory matrix gains exactly one count and loses a fraction $\alpha$ of counts per step, it would stabilize at an average count number $1/\alpha$ and lose \emph{on average} one memory per step. We encode the memories forgotten at step $t$ in the list of pairs $f(t)$ which is typically short (its length is a Poisson number with an average of 1). Given the random walk trajectory $x(t),x'(t)$ and the forgetting sequence $f(t)$, the memory matrix at any time point can be deterministically reconstructed as follows:
\begin{align}
	M_{ij}(t)=\sum\limits_{t'=1}^{t}\left( [i=x'(t')][j=x(t')] 
	-\sum\limits_{i,j\in f(t)}[i][j]\right),
	\label{eqn:reconalpha}
\end{align}
where the inner sum runs over the pairs stored in $f(t)$.

\subsection{The $\beta$ effect}
In the presence of the $\beta$ effect the random walk proceeds identically, but the memories are formed with a shuffling of the perceived step origin following the distribution $p(\Delta t)$ given by Eqn.~\ref{eqn:deltatdistr}. At every time step we draw a pseudorandom realization $p(\Delta t)\to \Delta t$, and update the stored trajectories as follows:
\begin{align}
	\left.P(j|i)\right|_{i=x(t)}\to x(t+1);\quad x'(t+1)=x(t-\Delta t),
\end{align}
which allows a deterministic reconstruction of the memory matrix with Eqn.~\ref{eqn:reconalpha}.

\subsection{The $\gamma$ effect}
In the presence of the $\gamma$ effect we need to compute not only a row of the textbook-based transition probability $T_{ij}(\tau)$, but also a row of the mental model $\hat{T}_{ij}(t)$ following Eqn.~\ref{eqn:mentalmodel}. If there are no memories in a given row (which for example is always the case on the very first random walk step), the mental model is zero and we use the textbook transition probability $P(j|i)=T_{ij}$. If there are memories, we compute the mixture of the two transition matrices with Eqn.~\ref{eqn:reinforcement}, use that to draw a pseudorandom realization of the next step $x(t+1)$, and proceed as before. The $\gamma$ effect easily combines with the $\alpha$ and $\beta$ effects. Note that the textbook transition probability $T_{ij}$ is always well defined for all nodes $i$ reachable through a random walk: if an edge led to the node, there is always at least that edge along which the random walker can return.

\section{Mental model metrics}
\label{app:precrec}
\subsection{Node metrics}
It is relatively straightforward to keep track of the number of nodes. The full network has $n$ nodes. By a specific sentence $\tau$ only a part of those nodes have been presented by the textbook, which we count as the number of rows in the adjacency matrix $\mathbf{A}(\tau)$ with non-zero entries:
\begin{align}
    n(\tau)=\sum\limits_i \left[ \left(\sum\limits_j A_{ij}(\tau)\right)>0 \right].
\end{align}

In a similar fashion we can count the number of nodes learned by the random walker, using either the memory matrix $\mathbf{M}$ or the normalized mental model $\hat{\mathbf{T}}$:
\begin{align}
    \hat{n}(t)=&\sum\limits_i \left[ \left(\sum\limits_j M_{ij}(t)\right)>0 \right] \nonumber \\
    =& \sum\limits_i \left[ \left(\sum\limits_j \hat{T}_{ij}(t)\right)>0 \right],
\end{align}
where the two definitions are equal because the absolute value of matrix elements does not matter. All that matters is their presence in rows. From the number of learned nodes we compute the node recall:
\begin{align}
    R_\textsf{n}=\hat{n}(t)/n,
\end{align}
where we divide by the total rather than the current number of nodes by convention. In this convention, the random walk starts with node recall 0 and can monotonically grow up to 1.

\subsection{Edge metrics}
The goal of constructing a mental model of network transitions is to predict the probability of transitioning from a given node $i$ to different nodes $j$. The absolute number of such transitions in either the textbook or the learner's memory should not matter. Therefore, in order to assess the quality of learning we seek a quantitative comparison mechanism between the taught transition matrix $\mathbf{T}$ and the learned one $\mathbf{\hat{T}}$. The relationship between them is illustrated by the Venn diagram in Fig.~\ref{fig:setup}f: generally, the two networks have partial overlap. Because of the interplay of finite learner effort and memory effects, the learned matrix might include spurious edges that were never taught, but lack taught edges that were never learned.

There are multiple ways to construct such comparison metrics. One way to compare the two networks is to treat them as conditional probability distributions and compute the Kullback-Leibler (KL) divergence between them as in Ref.~\cite{lynn2020human}. The KL divergence is zero when the two networks are identical and grows as the probability leaks into the spurious edges. However, a single missing taught edge immediately renders the KL divergence singular since it introduces a $\log(0)$ term into the sum. In order to avoid the divergence, we seek well-behaved metrics of the following form:
\begin{align}
    \textsf{Metric}=
    \frac{\sum_{ij}\textsf{Model}_{ij}[\textsf{Mask}_{ij}>0]}{\textsf{Norm}},
\end{align}
where each of the three components is a binary choice, resulting in eight possible metrics. The $\textsf{Model}$ component focuses on either learned or taught edges; the $\textsf{Mask}$ component focuses on either current or future edges; and the $\textsf{Norm}$ component focuses on either the complete network or the fraction of edges taught by a given time $\tau$. We use only a few of all possible component combinations, as detailed below.

As a first example, the edge recall $R_\textsf{e}$ is the fraction of edges of the taught mental model that have been learned (Fig.~\ref{fig:alpha1},\ref{fig:alpha3}):
\begin{align}
    R_\textsf{e}=\frac{1}{n}\sum\limits_{ij} T_{ij}[\hat{T}_{ij}>0],
    \label{eqn:edgerecall}
\end{align}
where the weight of each edge is given by the transition probability of the full textbook network; its inclusion is driven by the learned mental model; and the normalization equals the total number rows in either matrix, or the number of nodes in the network. As more and more edges are introduced in the taught network over time, more can be learned, thus increasing the edge recall metric. If all real edges have been learned, then the indicator function evaluates to 1 for all edges, and thus edge recall reaches 1.

In order to assess the precision of the learned mental model, we flip the taught and learned networks (Fig.~\ref{fig:beta1},\ref{fig:beta2}):
\begin{align}
    P=\frac{1}{n(\tau)}\sum\limits_{ij} \hat{T}_{ij}[T_{ij}>0],
\end{align}
where we also changed the normalization to refer to the nodes already introduced. If we used the fixed normalization $n$, then the magnitude of the precision metric would mostly follow the fraction of nodes learned: in other words, the number of rows with nonzero entries in $\hat{\mathbf{T}}$ would matter, rather than the content of those rows. Without shuffling ($\beta=\infty$), all learned edges necessarily exist so that $T_{ij}>0$ for any learned $(i,j)$. The precision can still be less than 1 since some rows of $\hat{\mathbf{T}}$ can still be empty due to under-sampling, and an absent mental model for transitions out of one node cannot be precise. If a learner's effort is sufficient to keep up with the introduction of new nodes $n(\tau)$, then a precision of 1 can be reached.

For measuring the prediction of future edges, we use the $\textsf{Mask}$ to select those (Fig.~\ref{fig:beta2}):
\begin{align}
    P^{pr}=&\frac{1}{\hat{n}(t)}\sum\limits_{ij} \hat{T}_{ij}[F_{ij}>\tau]\\
    R^{pr}=&\frac{\sum_{ij} T_{ij}[F_{ij}>\tau][\hat{T}_{ij}>\tau]}{\sum_{ij} T_{ij}[F_{ij}>\tau]},
\end{align}
where $[F_{ij}>\tau]$ picks out the real edges that would appear later than the current sentence $\tau$. For precision of prediction, we normalize by the number of nodes already learned by the random walker since any inferred connections are between those nodes. For the recall of prediction, we divide the total weight in the learned future edges by the total weight in all future edges. Both the numerator and the denominator of that expression approach zero by the end of the book, but the numerator is always no larger than the denominator, and hence the expression is never singular.

\subsection{Triangular closure}
Selecting the edges that comprise triangular closure of existing edges is another choice of $\textsf{Mask}$. A triangular closure is a walk of length 2 such that there is no direct edge between start and end. We therefore define a mask that is a product of those two conditions:
\begin{align}
    \textsf{TriClo}=[(\mathbf{A(\tau)}^2)_{ij}>0]\cdot[A_{ij}(\tau)=0],
\end{align}
and use this mask to select edges for scatter plots in Fig.~\ref{fig:beta2}.

\section{Exposure theory}
\label{app:exposure}
\begin{figure*}
	\includegraphics[width=0.7\textwidth]{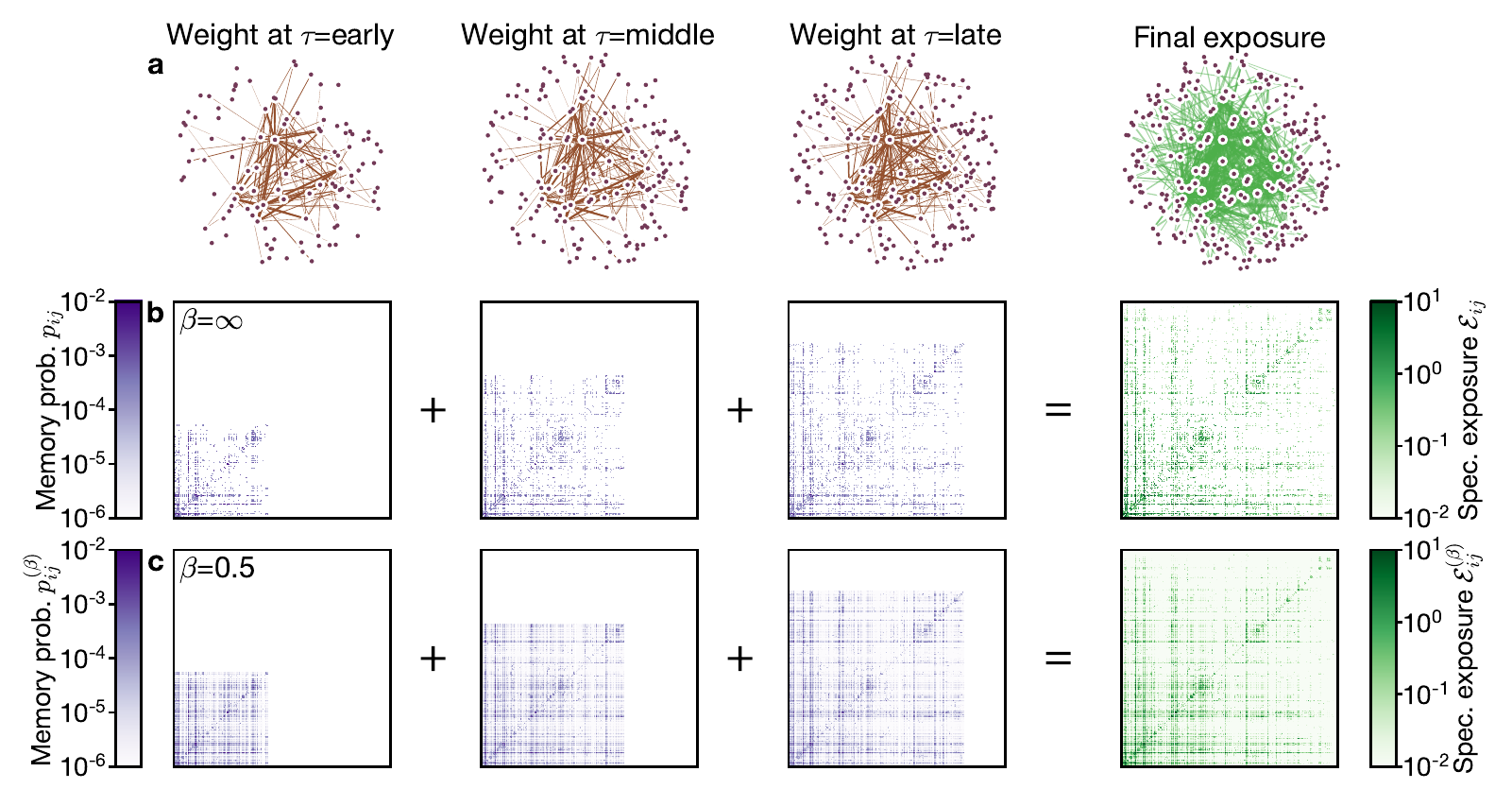}
	\caption{\textbf{The specific exposure matrix results from accumulation of memory probability.} (a) Textbook concept co-occurrence networks at early, middle, and late points in the textbook (brown), as well as a network with edges weighted by accumulated specific exposure (green). (b) As the network grows during exposition, the memory probability matrices $\mathbf{p}$ have more and more nonzero entries (purple). Their accumulation results in the specific exposure matrix $\mathcal{E}$ (green). (c) With finite shuffling ($\beta\neq \infty$) the memory probability matrices $\mathbf{p}^{(\beta)}$ become smudged across rows and columns (purple). The accumulation of shuffled memories results in the shuffled specific exposure $\mathbf{E}^{(\beta)}$.}
	\label{fig:cartoon}
\end{figure*}
\subsection{Baseline exposure}
The goal of exposure theory is to provide a computationally cheap but accurate approximation to the probability distribution of memory matrices $\mathbf{M}$, which in turn allows us to predict the trajectories of the mental model metrics. We first recap the baseline formulation of exposure theory as derived and validated in Ref.~\cite{klishin2022exposure}. We consider a weighted, undirected, time-dependent network described by the adjacency matrix $\mathbf{A}(\tau)$ (Fig.~\ref{fig:cartoon}). We assume that the network always has one main connected component and any disconnected pieces are small and only appear for a short time. The structure of the network is driven by the evolution time $\tau$, while the dynamics of the random walk are driven by the random walk time $t=D\tau$, where $D$ is the dilation parameter.

First we compute the steady-state probability of visiting a particular edge of the network. The transition matrix for the random walk is computed by normalizing the adjacency matrix by row sum, while the steady-state probability of each node is proportional to its strength (weighted degree):
\begin{align}
	T_{ij}(t)=\frac{A_{ij}(t)}{\sum_j A_{ij}(t)};\quad 
	\pi_i(t)=\frac{\sum_j A_{ij}(t)}{\sum_{ij} A_{ij}(t)},
\end{align}
and from these two expressions, we get the steady-state probability of visiting any edge of the network:
\begin{align}
	p_{ij}(t)=\pi_i T_{ij}(t)=\frac{A_{ij}(t)}{\sum_{ij}A_{ij}(t)}.
	\label{eqn:ptau}
\end{align}

The key assumption of exposure theory is that the random walk is always equilibrated to the instantaneous distribution. Practically, this happens when the correlation time of the random walk is much smaller than its length---an assumption that holds for many real-world networks (see Ref.~\cite{klishin2022exposure} for discussion). In this case, the accumulation of memory counts of any particular transition $M_{ij}$ is a Poisson process with the rate per step given by Eqn.~\ref{eqn:ptau}. For an equilibrated random walk the rate accumulates additively. The accumulation of the rate over time $t$ is termed the integral \emph{exposure}:
\begin{align}
	E_{ij}(t)\equiv 
	\sum\limits_{1}^{t}p_{ij}(\tau)=D\int\limits_{0}^{\tau}p_{ij}(\tau')d\tau'= 
	D\mathcal{E}_{ij}(\tau),
	\label{eqn:exposure}
\end{align}
where we changed variables between evolution time $\tau$ and random walk time $t$. The integral in $\tau'$, termed the \emph{specific exposure} $\mathcal{E}_{ij}$, can be precomputed at a desired time resolution with any standard method (Fig.~\ref{fig:cartoon}b). Converting from specific to integral exposure only requires a computationally cheap multiplication by dilation $D$. Once the integral exposure is known, the number of memories of the transition follows the Poisson distribution:
\begin{align}
	M_{ij}\sim \textsf{Pois}(D\mathcal{E}_{ij}(\tau)),
\end{align}
and specifically the probability that a transition has been seen at least once is given by:
\begin{align}
	P(M_{ij}>0)=1-e^{-D\mathcal{E}_{ij}(\tau)}.
	\label{eqn:probvisit}
\end{align}

Ref.~\cite{klishin2022exposure} also gives the rules of aggregation of exposure across a group of edges. For instance, from the edge exposure we can also compute the node exposure that accounts for the visitation of nodes:
\begin{align}
	K_i(t)\equiv \sum_j E_{ij}(t);\quad K_{i}(t)=D\mathcal{K}_i(\tau),
\end{align}
although the total exposure is conserved:
\begin{align}
    \sum\limits_i K_i = \sum\limits_{ij} E_{ij} = t = D\tau,
\end{align}
which sets the scale of memory fluctuations across the whole network.

In the limit of large dilation $D\to \infty$, the relative fluctuations in the memory counts get small and the Poisson random numbers are well-approximated by the mean, thus giving the \emph{mean-field limit}:
\begin{align}
	M_{ij}^\textsf{mf}\propto\mathcal{E}_{ij},
\end{align}
where the proportionality constant would cancel out from most computations of interest (e.g., row normalization).

\subsection{Node and edge recall}
In order to compute the exposure prediction of node and edge recall (Figs.~\ref{fig:alpha1},\ref{fig:alpha3}), we average the output of Eqn.~\ref{eqn:probvisit} over the node or edges of the network:
\begin{align}
    R_\textsf{n}=&1-\frac{1}{n}\sum\limits_i e^{-D\mathcal{K}_i(\tau)}\label{eqn:noderecexp}\\
    R_\textsf{e}=&1-\frac{1}{n}\sum\limits_{ij} T_{ij} e^{-D\mathcal{E}_{ij}(\tau)},\label{eqn:edgerecexp}
\end{align}
where we follow the weight convention of Appendix~\ref{app:precrec}.

\subsection{Jensen bound for nodes}
The shape of the node recall curve is subject to a Jensen bound (as derived in Ref.~\cite{klishin2022exposure}) by using the convexity property of $\phi(x)\equiv e^{-x}$:
\begin{align}
    R_\textsf{n}=1-\frac{1}{n}\sum\limits_i e^{-D\mathcal{K}_i(\tau)}\leq 1-e^{-t/n},
\end{align}
which implies that learning all nodes in a regular network (all nodes have the same strength) would be the fastest, with the timescale equal to the number of nodes $n$.

In practice we see that learning deeply under-saturates this bound (Figs.~\ref{fig:alpha1},\ref{fig:priority}). In order to explain this discrepancy, we expand the difference between the Jensen bound and the exposure prediction to second order in time $\tau$:
\begin{align}
    &R_\textsf{n}\simeq 1-\frac{1}{n}\sum\limits_i \left( 1-D\mathcal{K}_i(\tau) + \frac{1}{2}D^2 \mathcal{K}_i^2(\tau) + \order{\tau^3}\right)\nonumber\\
    &\quad= \frac{D\tau}{n} -\frac{D^2}{2n} \sum\limits_i \mathcal{K}_i^2(\tau)\\
    &R_\textsf{n}^{Jensen}\simeq  \frac{D\tau}{n}-\frac{D^2 \tau^2}{2n}+\order{\tau^3}\\
    &R_\textsf{n}^{Jensen}-R_\textsf{n} \simeq \frac{D^2}{2n}\sum\limits_i \mathcal{K}_i^2(\tau)-\frac{D^2 \tau^2}{2n}=\frac{D^2}{2}\textsf{Var}\;\mathcal{K}(\tau),
\end{align}
which directly connects the under-saturation of the Jensen bound with the variance (inhomogeneity, prioritization) of network nodes by specific exposure. The under-saturation only appears at second order in time, which explains why the tangents of the exposure curve and its bound coincide at the start (Figs.~\ref{fig:alpha1},\ref{fig:priority}).

\subsection{Jensen bound for edges}
The shape of the edge recall curve is subject to a similar bound that can be analogously derived:
\begin{align}
    R_\textsf{e}=1-\frac{1}{n}\sum\limits_{ij}T_{ij}e^{-D\mathcal{E}_{ij}(\tau)} \leq 1-\exp(-\frac{D}{n}\sum_{ij}T_{ij}\mathcal{E}_{ij}(\tau)),
\end{align}
where instead of equally-weighted average specific exposure over the nodes we now have a weighted average over the edges. The weights $T_{ij}$ stay constant in time, while the relative proportion of specific exposure on different edges shifts, so the weighted sum does not have a simple closed-form expression. However, we can approximate it: a typical transition probability out of a node equals either zero or its inverse degree, which we approximate by the inverse average degree $T_{ij}\approx n/m$. In this case the Jensen bound for edge recall takes the shape:
\begin{align}
    R_\textsf{e}\lesssim 1-e^{-t/m},
\end{align}
which predicts that an unweighted network would be the fastest to learn with a timescale equal to the number of edges $m$. Whereas at small $t$ this is not a strict bound (Figs.~\ref{fig:alpha1},\ref{fig:priority}), at larger $t$ the recall of weighted networks is also deeply unsaturated. This slowdown can be connected to the variance of edge exposure following a similar argument as for the nodes.

\subsection{The $\alpha$ effect}
In order to account for the $\alpha$ effect, we need to introduce forgetting into the memory dynamics. Since forgetting is stochastic but unbiased (every memory has an equal chance of being forgotten), we can just directly modify the exposure dynamics. The full stochastic process of memory dynamics is defined by master equations that are solved by an ansatz of a Poisson distribution with the to-be-determined parameter of exposure \cite{klishin2022exposure}. For normal learning, the exposure becomes a time integral of visitation probability (Eqn.~\ref{eqn:exposure}). The master equations with forgetting are still solved by a Poisson distribution ansatz, but with different dynamics of the exposure parameter. Across one time step, the exposure changes as follows:
\begin{align}
	E_{ij}(t+1)=E_{ij}(t)(1-\alpha)+p_{ij}(\tau),
	\label{eqn:Ealpha}
\end{align}
where $\alpha$ is the forgetting rate. This recursion relation can be solved by inductively substituting it into itself:
\begin{equation}
	E_{ij}(t)=\sum\limits_{t'=1}^{t}p_{ij}(\tau)(1-\alpha)^{t-t'}\simeq 
	D\int\limits_{0}^\tau p_{ij}(\tau) e^{-\alpha D (\tau-\tau')}d\tau',
\end{equation}
where we used $\alpha\ll 1$ and $t=D\tau$. Note that the difficult part of this expression (the integral) only depends on the \emph{product} of $\alpha D$ rather than on the two values individually. Thus computing it for a variety of $\alpha$ and $D$ value combinations only requires us to account for the distinct values that their product can take (Fig.~\ref{fig:alpha3}). In practice, the simplest way to compute the integral is to turn Eqn.~\ref{eqn:Ealpha} into an ordinary differential equation in $\tau$ and integrate it numerically following the scheme:
\begin{align}
	\mathcal{E}_{ij}^{(\alpha D)}(\tau+\Delta \tau)=\mathcal{E}_{ij}^{(\alpha 
	D)}(\tau)e^{-\alpha D\Delta \tau}+p_{ij}(\tau)\Delta \tau,
\end{align}
where $\Delta \tau$ is a suitably small integration step. Once the specific exposure is known, we convert it to the integral exposure $E_{ij}^{(\alpha D)}=D\mathcal{E}_{ij}^{(\alpha D)}$, which parameterizes the new Poisson distributions that now include forgetting. From those Poisson distributions, we compute the desired metrics in Fig.~\ref{fig:alpha3}.

\subsection{The $\beta$ effect: exposure accumulation}
In order to account for the $\beta$ effect of memory shuffling, we adapt the model of Ref.~\cite{lynn2020errors}. As the learner experiences a random walk, they do not remember the transitions exactly, but rather shuffle them locally. The shuffling distribution can be obtained from the free energy principle and has the geometric form $p(\Delta t)=(1-e^{-\beta})e^{-\beta\Delta t}$. In this case we are not interested only in the probability that a given edge $p_{ij}$ was visited but in the probability that a given edge $\hat{p}_{ij}$ was \emph{remembered}, after accounting for the memory shuffling. We know that the underlying random walk is the same and the probability of visiting any node $\pi_i$ is still the same. The remembered matrix of transitions after the shuffling was shown to be \cite{lynn2020errors}:
\begin{align}
	\mathbf{T}^{(\beta)}=(1-e^{-\beta})\sum\limits_{k=0}^\infty e^{-\beta 
	k}\mathbf{T}^{k+1}=(1-e^{-\beta})\mathbf{T} 
	\left(\mathbf{I}-e^{-\beta}\mathbf{T}\right)^{-1}.
\end{align}

The original and shuffled transition matrices $\mathbf{T}$ and $\mathbf{T}^{(\beta)}$ are both row-normalized to describe the learned transition rates. In order to find the absolute, rather than conditional, probability of learning a particular edge, we multiply it by $\pi_i$:
\begin{align}
	p^{(\beta)}_{ij}(t)=&\pi_i(t) T^{(\beta)}_{ij}(t)\\
	\lim\limits_{\beta\to 0}p^{(\beta)}_{ij}(t)=&\pi_i(t) \pi_j(t),
\end{align}
where the matrix $\mathbf{p}^{(\beta)}$ can be checked to be symmetric and normalized so that all entries sum up to 1. In the complete shuffling limit $\beta\to 0$, the learner loses all notion of the order of explored nodes, but still keeps track of the relative frequency of visiting different nodes.

We next take the same assumption as in the first derivation of exposure theory: that the random walk is always equilibrated, now also with respect to memory shuffling. At finite $\beta$, the transition memories are shuffled on the timescale of roughly $\expval{\Delta t}\simeq 1/(e^\beta-1)$. We assume that this timescale, just like the random walk correlation time, is much shorter than the timescale of network exploration. Therefore, at each step of the network, memories of each edge (real or spurious) are created with probabilities $p^{(\beta)}_{ij}$. Note that the spurious edges can only be generated between the nodes that have already been introduced. The overall specific exposure can be computed by integrating the remembering probability (Fig.~\ref{fig:cartoon}c):
\begin{align}
	\mathcal{E}_{ij}^{(\beta)}(\tau)= 
	\int\limits_{\tau'=0}^{\tau}p^{(\beta)}_{ij}(\tau')d\tau',
\end{align}
which for $\beta\to \infty$ reduces to the old formula (Eqn.~\ref{eqn:exposure}). The integral exposure is obtained, just like before, by rescaling the specific exposure by dilation $E_{ij}^{(\beta)}=D\mathcal{E}_{ij}^{(\beta)}$. Since some of the exposure now falls onto the spurious edges, the exposure of real edges is necessarily smaller, and thus it would take longer to learn the real edges. As before, the prediction of edge learning is more accurate in the aggregate, which can now be extended to the spurious edges.

\subsection{The $\beta$ effect: metric computation}
First we compute the precision in learning the real edges, i.e., the proportion of probability weight in the learned mental model that lies in real edges. With exposure theory we can evaluate the two contributions to this precision: whether each node has been visited at all and what fraction of memories transitioning out of that node corresponds to real edges. By a certain time $\tau$, $n(\tau)$ nodes of the network have been introduced. The specific exposure of every node is given by:
\begin{align}
    \mathcal{K}_i(\tau)=\sum\limits_j \mathcal{E}^{(\beta)}_{ij},
\end{align}
and is independent of $\beta$ since regardless of shuffling the random walker always keeps track of the node they just arrived on. For the nodes that were not introduced yet, $\mathcal{K}_i=0\;\forall i>N(\tau)$. The combination of specific exposure and dilation predicts the probability of the node being visited.

If a node has been visited, we can define two groups of outgoing edges: edges that are real and all edges (the first is a subset of the second). Since the accumulation of all edge memories is independent, we can use Poisson calculus to compute the number of memories in each group by selecting the real edges with a $\textsf{Mask}$ \cite{klishin2022exposure}:
\begin{align}
    M_i^\textsf{real}\sim&\textsf{Pois}\left(\sum\limits_j D\mathcal{E}_{ij}^{(\beta)}[A_{ij}>0] \right)\\
    M_i^\textsf{all}\sim&\textsf{Pois}\left(\sum\limits_j D\mathcal{E}_{ij}^{(\beta)} \right),
\end{align}
from which we can estimate the fraction of weight in the real edges. Putting the node and edge contributions together, we get the following exposure prediction of precision:
\begin{align}
	P(\tau)=\frac{1}{n(\tau)}\sum_i (1-e^{-D\mathcal{K}_i})\frac{\sum_j 
	\mathcal{E}_{ij}^{(\beta)} [A_{ij}>0]}{\sum_j \mathcal{E}_{ij}^{(\beta)}},
\end{align}
where dilation cancels out from the final fraction: as dilation gets larger, the counts of real and spurious edges follow the same proportion. The precision metric only depends on dilation through visitation of nodes. As $D\to \infty$, precision approaches a finite value dependent on $\tau$ and $\beta$ (Fig.~\ref{fig:beta1},\ref{fig:beta2}).

We can similarly compute the precision of edge prediction, i.e., the fraction of probability weight in edges that will appear later (Fig.~\ref{fig:beta2}). This calculation only requires swapping out the mask in the numerator from edge existence $[A_{ij}>0]$ to filtration order $[F_{ij}>\tau]$:
\begin{align}
	P^{pr}(\tau)=\frac{1}{n(\tau)}\sum_i (1-e^{-D\mathcal{K}_i})\frac{\sum_j 
		\mathcal{E}_{ij}^{(\beta)} [F_{ij}>\tau]}{\sum_j 
		\mathcal{E}_{ij}^{(\beta)}},
\end{align}
where by the end of the book there are no edges left such that $F_{ij}>\tau$, and thus any prediction is impossible.

In order to compute the recall of prediction (Fig.~\ref{fig:beta2}), we consider the transition matrix of the whole book that ought to be learned $T_{ij}$, filter the edges that would exist in the future, and account for the probability of those edges being remembered:
\begin{align}
	R^{pr}(\tau)=\frac{\sum_{ij} T_{ij} (1-e^{-D\mathcal{E}_{ij}}) 
	[F_{ij}>\tau]}{\sum_{ij} T_{ij} [F_{ij}>\tau]}.
\end{align}

\subsection{The $\gamma$ effect}
In order to account for the $\gamma$ effect, we need to provide two exposure-based computations: the overlap metric and the memory count asymmetry. The presence of reinforcement in general breaks the ergodicity of memory accumulation dynamics and thus breaks the core assumption of exposure theory. So instead, we compute the exposure theory predictions \emph{in the absence} of reinforcement $\gamma=0$, and show how they can relate to a finite $\gamma$ case.

The overlap metric (Eqn.~\ref{eqn:overlap}) is quite similar to edge recall (Eqn.~\ref{eqn:edgerecall}) but involves an unweighted average over the edges. We can therefore construct an exposure prediction similar to Eqn.~\ref{eqn:edgerecexp}. Per exposure theory, the learning of each edge is independent from learning any other edge in the same replica of a random walk, and also independent from learning the same edge in a different replica. Since the exposure metric can either compare a replica to itself (self-overlap $Q_{aa}$) or to another replica (cross-overlap $Q_{ab}$), we construct the following two estimators:
\begin{align}
    Q_{aa}(\tau)=&\frac{1}{m}\sum\limits_{ij}[T_{ij}>0]\left( 1-e^{-D\mathcal{E}_{ij}(\tau)} \right)\\
    Q_{ab}(\tau)=&\frac{1}{m}\sum\limits_{ij}[T_{ij}>0]\left( 1-e^{-D\mathcal{E}_{ij}(\tau)} \right)^2,
\end{align}
where we used the fact that the probability of two identically distributed independent events happening is the square of the probability of one event. Since the probabilities are less than or equal to 1, the terms in the cross-overlap sum are typically smaller than in the self-overlap sum. For self-overlap $Q_{aa}$ we can construct a Jensen bound by again using the concavity of the function $\phi(x)=(1-e^{-x})$. In contrast, for cross-overlap $Q_{ab}$, such a Jensen bound does not apply because the function $\phi(x)=(1-e^{-x})^2$ is neither convex nor concave and has an inflection point. While for a given network the above formulas make accurate predictions of self- and cross-overlap, it is harder to make general claims about the space of possible networks.

How does this prediction at $\gamma=0$ help us to reason about finite reinforcement $\gamma>0$? The key idea is that discovery of new edges is driven by independent random walk steps and not reinforcement. The accumulation of the first memory count is still a Poisson process predicted by Eqn.~\ref{eqn:probvisit}, even if the rest of the distribution is different due to reinforcement. However, in the presence of reinforcement not all steps are independent. If over the course of a long random walk a total of $D\tau$ steps have been made, of those roughly $D\tau (1-\gamma)$ steps followed the network and had a chance to discover new edges, and the other $D\tau \gamma$ steps retraced known edges. The independent steps are uniformly distributed among all steps, following the evolution of the network. Therefore in order to predict the self- and cross-overlap in the presence of reinforcement, we can pre-compute the curves $Q_{aa}(D)$ and $Q_{ab}(D)$, and look up the value at the effective dilation $D_\gamma=D(1-\gamma)$ (Fig.~\ref{fig:gamma1}).

The prediction of memory asymmetry proceeds similarly: we estimate the asymmetry in the absence of reinforcement $\gamma=0$ and determine whether the estimate is broken for $\gamma>0$. Without reinforcement, the number of memories of each edge is a non-negative integer with a Poisson distribution parameterized by the integral exposure:
\begin{align}
    M_{ij}&\in \{ \mathbb{Z}\geq 0 \}\\
    M_{ij}&\sim\textsf{Pois}(D\mathcal{E}_{ij}(\tau)),
\end{align}
and the distributions for the reciprocal edge $M_{ji}$ are identical. The difference of the two can be any integer and follows the Skellam distribution:
\begin{align}
    M_{ij}-M_{ji}&\in \mathbb{Z}\\
    M_{ij}-M_{ji}&\sim \textsf{Skellam}(D\mathcal{E}_{ij}(\tau),D\mathcal{E}_{ij}(\tau)),
\end{align}
where the exact functional form of the distribution can be computed but is not very important as we instead focus on its moments. When two independent random numbers are subtracted, their means subtract but their variances add:
\begin{align}
    \expval{M_{ij}-M_{ji}}=&0\\
    \expval{(M_{ij}-M_{ji})^2}_c=&2D\mathcal{E}_{ij}(\tau),
    \label{eqn:asymmvariance}
\end{align}
and the standard deviation of asymmetry is the square root of variance $\sqrt{2D\mathcal{E}_{ij}(\tau)}$. This standard deviation defines the expected range of asymmetry, which we check for violations in the presence of reinforcement (Fig.~\ref{fig:gamma2}).

\section{Supplementary results}
\label{app:supp}
\begin{figure*}
	\includegraphics[width=\textwidth]{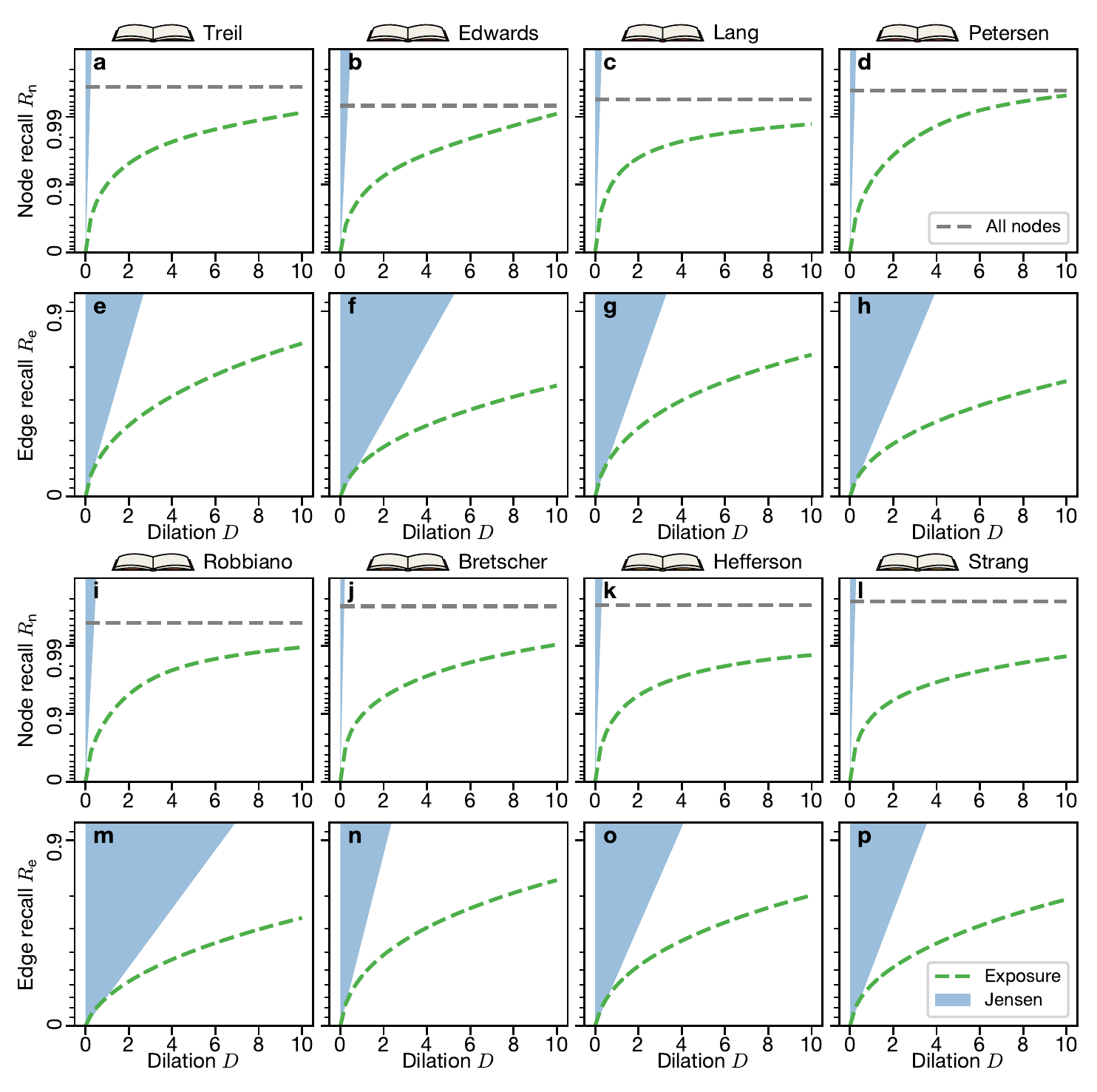}
	\caption{\textbf{Comparison of node and edge recall trajectories for the remaining eight textbooks.} For each textbook, on the top plot (panels a-d, i-l) the green dashed curve shows the exposure prediction of node recall; the blue shaded region shows the Jensen bound; and the gray dashed horizontal line shows the learning of all but one node ($R_\textsf{n}=1-1/n$). On the bottom plot (panels e-h, m-p) the green dashed curve shows the exposure prediction of edge recall, and the blue shaded region shows the Jensen bound.}
	\label{fig:priority}
\end{figure*}

\begin{figure*}
	\includegraphics[width=\textwidth]{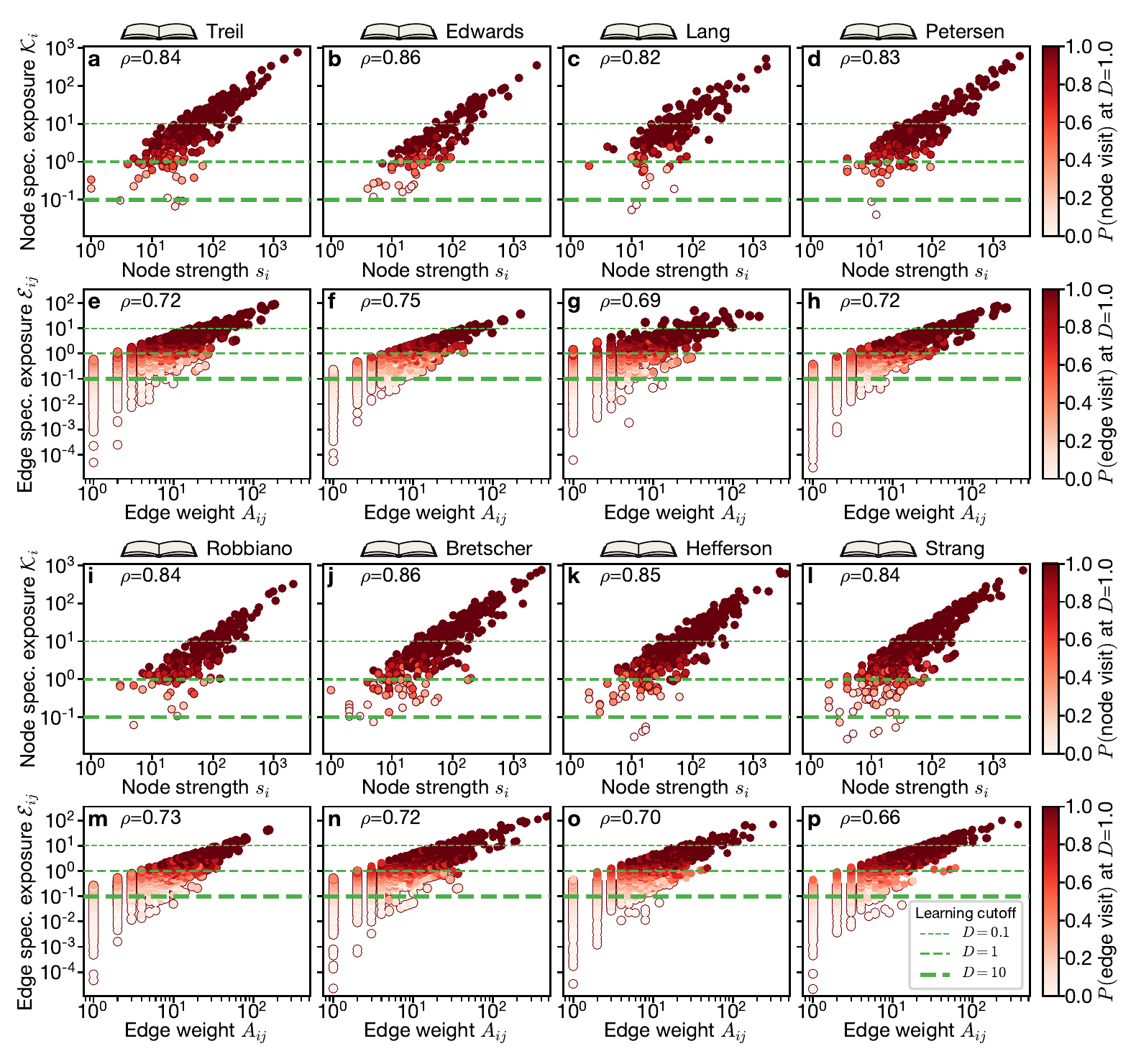}
	\caption{\textbf{Comparison of node and edge exposure pattern for the remaining eight textbooks.} For each textbook, the top scatter plot shows the node specific exposure and strength, whereas the bottom scatter plot shows the edge specific exposure and edge weight. The marker color corresponds to the probability of a node or edge being learned across 10 replicas at $D=1.0$. The horizontal dashed lines indicate the boundary of node and edge learning at different dilation. In each panel $\rho$ is the Spearman correlation coefficient, with $\log_{10}(p)<-12$.}
	\label{fig:comparison}
\end{figure*}

\begin{figure*}
	\includegraphics[width=\textwidth]{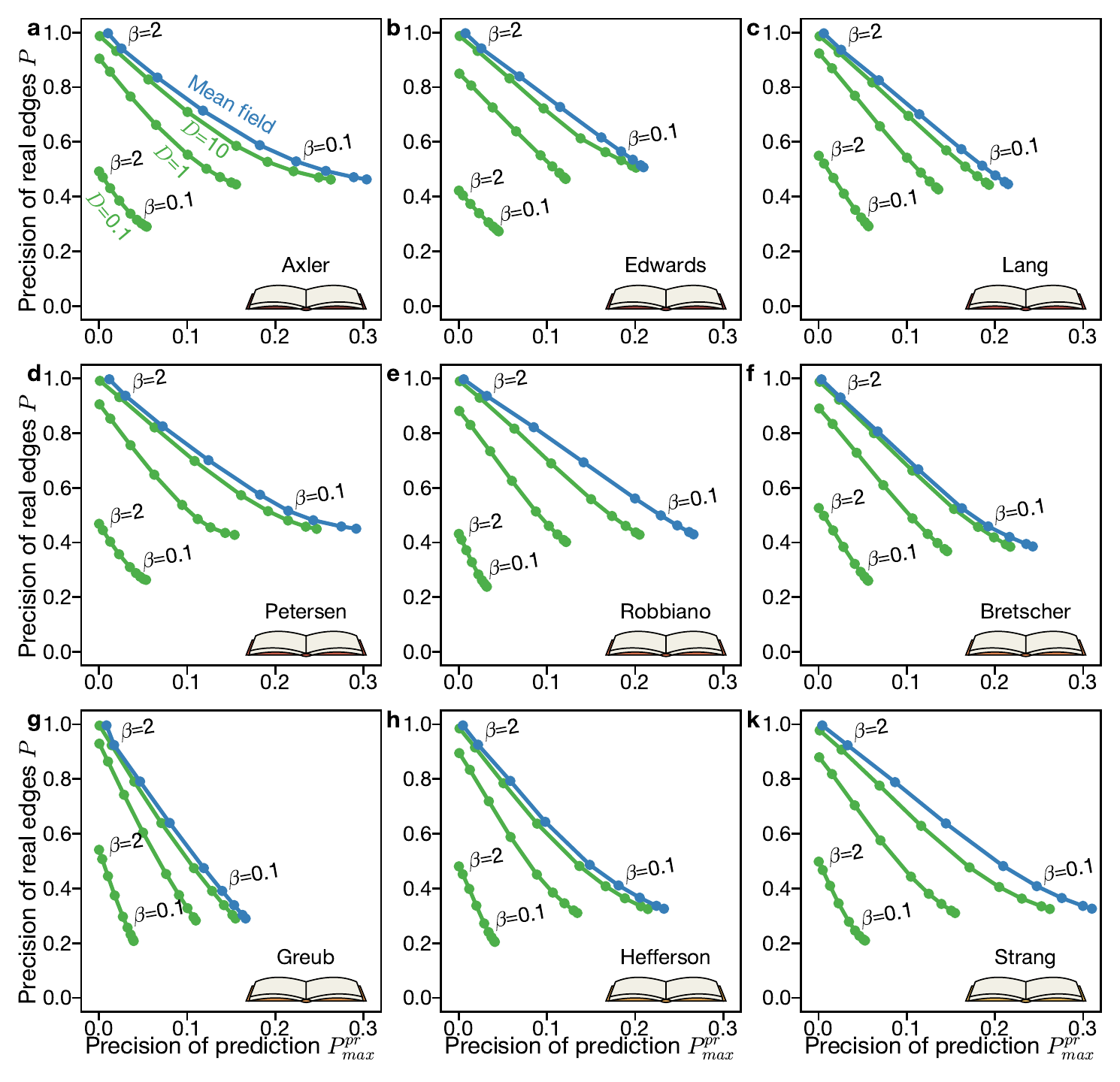}
	\caption{\textbf{Comparison of the precision-prediction trade-off for the remaining nine textbooks.} For each textbook, each curve shows variable shuffling $\beta$, while different curves correspond to different dilation $D$.}
	\label{fig:tradeoff}
\end{figure*}

In the main text of the paper we explored the consequences of finite effort and memory effects on network learning for a few example textbooks. Here we provide identical analyses for the rest of the textbooks.

We previously showed the comparison of typical learning trajectories for two textbooks (Fig.~\ref{fig:alpha1}). The other eight textbooks show qualitatively similar curves (Fig.~\ref{fig:priority}), confirming that prioritization of concepts and connections is generically present across most textbooks. All of the recall trajectories are deeply unsaturated compared to the respective Jensen bounds. In the plotted dilation range $D\in[0,10]$, none of the textbooks reliably reach the learning of all but one node $R_\textsf{n}=1-1/n$.

We explained the global learning trajectories in terms of local specific exposure of nodes and edges for two textbooks (Fig.~\ref{fig:alpha2}). For the other textbooks, the patterns are qualitatively similar (Fig.~\ref{fig:comparison}).

We compared the precision of learning real edges with the peak precision of predicting future edges for a single textbook (Fig.~\ref{fig:beta2}). Across other textbooks, the trade-off pattern is broadly similar (Fig.~\ref{fig:tradeoff}). The lowest precision of real edges varies in the range $[0.3,0.5]$, while the highest prediction of future edges varies in the range $[0.17,0.31]$. Across all textbooks, the finite values of dilation $D$ significantly limit both precision metrics.

\clearpage
\newpage
\bibliography{biblio}

\end{document}